\documentclass[prx,twocolumn,english,superscriptaddress,floatfix,longbibliography]{revtex4-2}

\usepackage{amsmath}
\usepackage{amssymb}
\usepackage{stmaryrd}
\usepackage{array}
\usepackage{tabularx}
\usepackage{booktabs}
\usepackage{placeins}
\usepackage{svg}
\usepackage[separate-uncertainty=true]{siunitx}
\definecolor{bluepoli}{RGB}{0,36,179}
\definecolor{redpoli}{RGB}{204,0,51}
\definecolor{greenpoli}{RGB}{45,137,0}
\definecolor{purplepoli}{RGB}{153,102,204}
\definecolor{azzurropoli}{RGB}{51,53,204}
\definecolor{orangepoli}{RGB}{255,124,17}
\usepackage{booktabs}
\definecolor{rowshade}{gray}{0.92}
\usepackage{hyperref}
\usepackage{physics}
\usepackage{upgreek} 
\hypersetup{%
	colorlinks,
	linkcolor={bluepoli}, %
	citecolor={redpoli}, %
	urlcolor={redpoli} %
}
\usepackage{algpseudocode}  
\usepackage{etoolbox}       
\usepackage[ruled,vlined,linesnumbered]{algorithm2e}
\usepackage[separate-uncertainty=true]{siunitx}

\usepackage{makecell}  
\usepackage{multirow}

\def\RLEaffil{Research Laboratory of Electronics, Massachusetts Institute of Technology, Cambridge, MA 02139, USA}
\def\Physaffil{Department of Physics, Massachusetts Institute of Technology, Cambridge, MA 02139, USA}
\def\EECSaffil{Department of Electrical Engineering and Computer Science, Massachusetts Institute of Technology, Cambridge, MA 02139, USA}

\def\ETHaffil{Department of Information Technology and Electrical Engineering, ETH Zürich, ZH 8092, Switzerland}

\begin{document}

    \title{Placing and routing quantum LDPC codes in multilayer superconducting hardware }

    \author{Melvin~Mathews}
        \altaffiliation{These authors contributed equally to this work. \\
        MM present address: melvinmathews@google.com}
        \affiliation{\RLEaffil}
        \affiliation{\ETHaffil}
    \author{Lukas~Pahl}
         \altaffiliation{These authors contributed equally to this work. \\
        MM present address: melvinmathews@google.com}
        \affiliation{\RLEaffil}
        \affiliation{\EECSaffil}
    \author{David~Pahl}
         \altaffiliation{These authors contributed equally to this work. \\
        MM present address: melvinmathews@google.com}
        \affiliation{\RLEaffil}
        \affiliation{\EECSaffil}
    \author{Vaishnavi~L.~Addala}
        \affiliation{\RLEaffil}
        \affiliation{\EECSaffil}
    \author{Catherine~Tang}
        \affiliation{\RLEaffil}
        \affiliation{\EECSaffil}
    \author{William~D.~Oliver}
        \affiliation{\RLEaffil}
        \affiliation{\EECSaffil}
        \affiliation{\Physaffil}
    \author{Jeffrey~A.~Grover}
        \email{jagrover@mit.edu}
        \affiliation{\RLEaffil}

    \date{\today}
    
    \begin{abstract}
       
       Quantum error-correcting codes with asymptotically lower overheads than the surface code require nonlocal connectivity~\cite{bravyi2010tradeoffs}. Leveraging multilayer routing and long-range coupling capabilities in superconducting qubit hardware, we develop Hardware-Aware Layout, HAL: a robust, runtime-efficient heuristic algorithm that automates and optimizes the placement and routing of arbitrary codes. Using HAL, we generate around 150 explicit layouts of quantum low-density parity-check (qLDPC) codes with topological structure---such as the bivariate bicycle codes~\cite{liang2025generalized, bravyi2024high} and the open-boundary tile codes~\cite{steffan2025tile, liang2025planar}---and find that removing the periodic boundaries significantly lowers the hardware complexity with only a moderate reduction of logical efficiency. We also lay out highly nonlocal qLDPC code families--- quantum radial~\cite{scruby2024high} and Tanner codes~\cite{radebold2025explicit}---that achieve competitive tradeoffs between hardware complexity and logical efficiency. Based on our findings, we anticipate many novel qLDPC codes to be realizable on near-term superconducting qubit hardware and inform future directions for the co-design of quantum devices and fault-tolerant architectures.

    \end{abstract}
    
\maketitle

    \section{Introduction}  
    
    Since its introduction, the surface code~\cite{bravyi1998quantum,freedman2001projective,fowler2012surface} remains one of the most promising quantum error-correcting codes (QECCs) for near-term, fault-tolerant devices. This is mainly due to its local and planar structure, making it particularly well suited for current solid-state quantum hardware, where connectivity is often limited~\cite{andersen2020repeated, krinner2022realizing, google2023suppressing, acharya2024quantum}. This locality comes at the cost of a large qubit overhead. In recent years, quantum low-density parity-check (qLDPC) codes have emerged as a promising alternative to the surface code~\cite{breuckmann2021quantum, bravyi2024high}. These codes achieve higher logical-qubit-encoding rates and better distance scaling than the surface code. However, qLDPC codes typically require long-range connectivity between qubits~\cite{bravyi2010tradeoffs}.

    While this nonlocality presents a challenge for superconducting hardware, significant advances in design and fabrication have raised hopes that some nonlocality will be realizable in superconducting systems. This includes the bump-bonding of chips~\cite{rosenberg20173d,field2024modular, kosen2024signal, karamlou2024probing, norris2024improved, norris2025performance} and the use of superconducting through-silicon vias (TSVs)~\cite{yost2020solid,mallek2021fabrication,hazard2023characterization}, which together allow for routing on multiple layers. Paired with the realization of long-range couplers~\cite{storz2023loophole, wang2025demonstration, kumph2024demonstration, niu2023low, marxer2023long, xiong2025scalable, heya2025randomized, xu2025tunable}, these technologies enable vertically integrated devices with more complex connectivities.
    
    As these hardware capabilities mature and more qLDPC codes are developed, it is important to ask: how readily can these codes be implemented on superconducting qubit architectures? Prior work has considered the use of gate teleportation or SWAP gates to measure nonlocal stabilizers. Such approaches minimize the number of required physical long-range couplers~\cite{delfosse2021bounds,berthusen2024toward,xiong2025scalable}. However, they can result in increased qubit numbers and circuit depths, incurring time overheads and potentially low to non-existent thresholds. These challenges can be partially mitigated at large scales using code concatenation~\cite{pattison2025hierarchical, gidney2025yoked}.

    An alternative is to realize the fixed physical connections required to measure all stabilizers directly. Recent work~\cite{tremblay2022constant, bravyi2024high} suggests partitioning the connectivity graphs of qLDPC codes into multiple layers of planar subgraphs. Since qubit positions are fixed across subgraphs, this approach can come at the cost of many parallel, long routed edges~\cite{pach1998embedding,schaefer2021new}, which may be impractical under realistic hardware constraints. Furthermore, Refs.~\cite{tremblay2022constant, bravyi2024high} do not contain explicit, geometric layouts of qLDPC codes or discuss an algorithmic method to place and route arbitrary QECCs on multilayer hardware.

    \begin{figure*}
        \centering
        \includegraphics[width = 2\columnwidth]{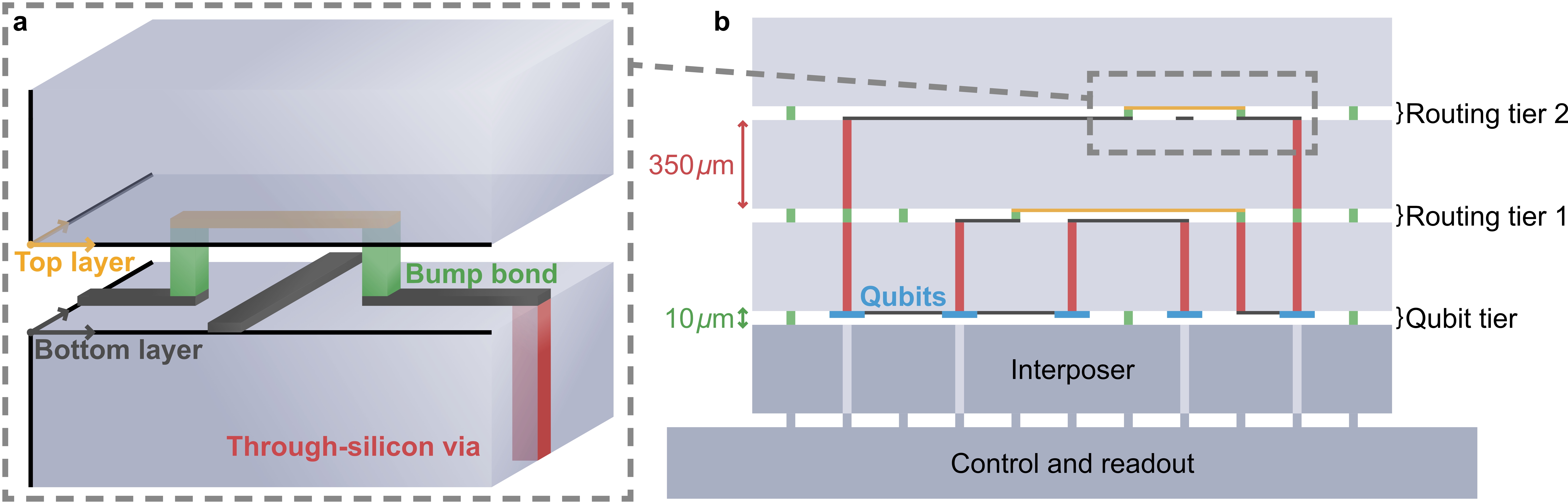}
        \caption{\textbf{Superconducting hardware stackup and definitions.} \textbf{a} The flip-chip geometry, in which two chips are mechanically bonded with bump bonds (green). In the resulting inter-chip space, signal lines can be routed on both the bottom and top layers, and even transition through the bump bonds. The spacing is typically on the order of $\SI{10}{\micro\meter}$. Superconducting through-silicon vias (TSVs) (red) are used to transition to other tiers. \textbf{b} Cross-sectional schematic of a proposed multilayer superconducting quantum architecture enabling complex connectivities. The stackup consists of a qubit chip stacked on an interposer and a superconducting multi-chip module, vertically connecting the qubits with control and readout circuitry. Both layers in the qubit tier are available for hosting qubits, couplers, and readout and control circuitry. In this work, we choose—without loss of generality—to place both qubits and couplers on the upper layer. Further chips are stacked on top of the qubit chip to realize coherent routing of long-range couplers. The edges traverse TSVs and bumps to reach higher tiers and routing layers. This architecture serves as the physical framework for HAL. Note that the TSVs and bump bonds beneath the qubit tier are colored in gray to indicate that they are not used to route long-range couplers.}
        \label{fig:stackup}
    \end{figure*}

    We address these limitations by taking advantage of near-term superconducting fabrication advances, such as the ability of an edge to transition between layers along its path, enabling shorter coupler lengths. Building on this, we introduce HAL (Hardware-Aware Layout)---a tool that automates and optimizes the placement and routing of connectivity graphs for arbitrary QECCs on multilayer hardware. Our work closely parallels the place-and-route problem in classical integrated-circuit design, where growing circuit complexity and advances in semiconductor fabrication drove the development of electronic design automation (EDA)~\cite{kahng2011vlsi}. Likewise, the emergence of multilayer superconducting circuits and rapidly evolving qLDPC codes calls for automated tools that integrate code layout with hardware constraints.

   The automated nature of HAL enables the extraction of architectural insights across many qLDPC code families. In this work, we study bivariate bicycle codes \cite{liang2025generalized}, open-boundary tile codes~\cite{steffan2025tile}, toroidally planar directional codes~\cite{geher2025directional}, constant-depth-decodable radial codes~\cite{scruby2024high}, and asymptotically good Tanner codes~\cite{radebold2025explicit}. Among several findings, we confirm a tradeoff between connectivity and logical efficiency. We also anticipate many qLDPC codes to be realizable in the near-term and point to directions to further improve their feasibility.

    The remainder of this paper is organized as follows. In Sec.~\ref{sec:stackup}, we discuss near-term superconducting hardware capabilities, with which we define the stackup assumed by HAL. Sec.~\ref{sec:algo} describes the core algorithmic workflow of HAL. In Sec.~\ref{sec:results}, we present explicit code layouts for a bivariate bicycle, tile, and radial code. In Sec.~\ref{sec:results}, we define a hardware-complexity metric and quantify this metric across our full set of code families.
    
    \section{Superconducting Stackup}
    \label{sec:stackup}
    
    With ongoing advances in fabrication techniques, superconducting qubit devices can begin to move beyond nearest-neighbor lattices. For instance, bump-bonding of two superconducting chips to implement the flip-chip geometry has been well-established for several years~\cite{rosenberg20173d,field2024modular, kosen2024signal, karamlou2024probing, norris2025performance}. In this geometry, depicted in Fig.~\ref{fig:stackup}\textbf{a}, two chips are mechanically bonded together through the use of indium bumps (green), forming two opposing layers with inter-chip spacings of \qtyrange[range-units=single,range-phrase=--]{3}{15}{\micro\meter}. Signals can be routed on both layers and even traverse between layers through the bump bonds. In Refs.~\cite{field2024modular, norris2025performance}, high-fidelity two-qubit gates were demonstrated with couplers that traversed through multiple bumps across the inter-chip layers of flip-chip devices. 
    
    Superconducting TSVs (red) provide vertical signal connections between the bottom and top sides of a chip. In Ref.~\cite{hazard2023characterization}, qubits with most of their capacitance coming from TSVs showed quality factors of around $750\times10^3$, indicating that couplers can remain coherent when routed through TSVs. Note that TSVs can also be used to compactly route all control and readout circuitry vertically to the qubits~\cite{yost2020solid}.

    The development of long-range, on-chip couplers is a topic of active investigation~\cite{wang2025demonstration, kumph2024demonstration, marxer2023long, xiong2025scalable,heya2025randomized}. There is clear evidence that high-fidelity two-qubit gates mediated by centimeters-long couplers will be possible in the near future.

    \begin{figure*}
        \centering
        \includegraphics[width=1.8\columnwidth]{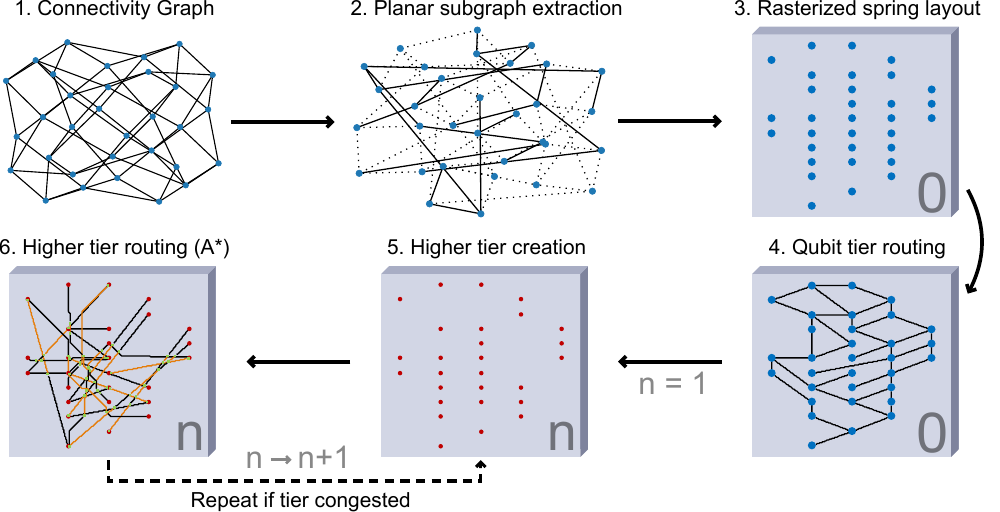}
        \caption{\textbf{The algorithmic workflow of HAL for laying out arbitrary quantum error-correcting codes.} The process begins with a (1) connectivity graph (shown here for a $\llbracket16,2,4\rrbracket$ radial code) and proceeds through four key stages: (2) heuristic maximum planar subgraph extraction, separating the graph into planar (solid) and non-planar (dashed) components; (3) node placement via a spring layout and rasterization; (4) placing the qubit tier edges; (5) higher tier creation with the necessary TSVs (red dots); it is implied that each route uses a dedicated TSV in the vicinity of the red dots to traverse vertically through tiers. (6) edge routing on this higher tier using a modified A* pathfinding algorithm, using bump transitions to resolve crossings. Steps 5 and 6 are repeated once a higher tier becomes congested. In the above case, HAL returns the tiers 0 and 1 for the $\llbracket16,2,4\rrbracket$ code.}
        \label{fig:algo}
    \end{figure*}

    Given these and future developments, we propose a multi-chip stackup for realizing elaborate connectivity, as depicted in Fig.~\ref{fig:stackup}\textbf{b}.  We closely follow Ref.~\cite{yost2020solid}, which proposes a qubit chip stacked onto an interposer and a superconducting multi-chip module. In this stackup, qubits are vertically connected to the readout and control circuitry below them. We suggest extending this architecture by stacking additional chips on top of the qubit chip to provide routing layers for long-range couplers. Each stacked chip introduces an additional routing tier comprising the two opposing signal layers, formed by the inter-chip space of the flip-chip geometry (Fig.~\ref{fig:stackup}\textbf{a}).
    
    Both layers in the qubit tier are available for qubits, couplers, and readout and control circuitry. Without loss of generality, in this work, we elect to place both qubits and couplers on the top layer of the qubit tier. In principle, qubits could also be added on higher routing tiers; we do not make this assumption in our work, but incorporating such qubits would be straightforward within our framework. Couplers that cannot be routed on the qubit tier without crossings can be vertically routed through a TSV to a higher tier and traverse to the location of their target qubit. Importantly, within each higher routing tier, both opposing layers can be used to route couplers, and couplers can transition between these layers. HAL relies on this bi-layer routing capability to resolve crossings in higher routing tiers.
    
    In realistic hardware, the crosstalk between two crossing lines on opposing layers within one tier can be suppressed with increasing inter-chip spacing and additional shielding methods such as enclosing the routes in metal tunnels~\cite{bu2025tantalum, kosen2024signal} as well as strategic placement of bump bonds. The crosstalk between lines on separate sides of a chip is suppressed due to their far separation and can be mitigated using further shielding techniques, such as intermediate ground planes and via shielding. Overall, this multilayer superconducting architecture improves connectivity over nearest-neighbor lattices, enabling the implementation of complex qLDPC codes.
    
    \section{HAL Algorithm}
    \label{sec:algo}
    To place and route a QECC, HAL first requires its connectivity graph, where the nodes represent all physical qubits, including data and check qubits (see Fig.~\ref{fig:algo}). The edges between data and check qubits represent the physical couplers required to implement the necessary parity checks natively. Note that, in this work, we infer the connectivity graph directly from the parity check matrix of a code. When placing and routing a connectivity graph onto multilayer superconducting hardware, we treat check and data qubits equally and safely ignore the Pauli basis of parity checks. We define the following constraints and desiderata:

    \begin{enumerate}
        \item \textbf{Qubits}: Qubits are placed on discrete grid points. This regularity simplifies the design and routing of control and readout lines and guarantees sufficient spacing between qubits.
        \item \textbf{Tiers}: A core objective is to minimize the total number of tiers. The qubit tier plays a special role as it hosts the qubits.
        \item \textbf{Couplers}: It is desired to have most edges on the qubit tier, with fewer edges on each higher tier. This distribution of edges minimizes the amount of TSVs used per coupler, keeping couplers coherent. Additionally, edges on the same layer cannot cross. To resolve such a crossing, the coupler representing the edge must either be routed around the intersecting coupler, moved to the opposing layer through a bump bond, or moved to another tier through a TSV. The length, number of bump bonds, and number of TSVs per coupler are ideally minimized.
    \end{enumerate}

    The algorithmic workflow of HAL is illustrated in Fig.~\ref{fig:algo}. A detailed description can be found in App.~\ref{App_algo}. Laying out a connectivity graph is generally subdivided into two parts: placing the nodes and routing the edges. We begin by describing the node placement step.

    If a code possesses a clear geometric structure, as is the case with 2D-topological codes, where the qubits are arranged in a square lattice, the user may provide qubit positions to HAL to enforce this structure.
    
    However, we also developed a generic placement algorithm to handle arbitrary connectivity graphs, even when they do not admit a clear geometric structure. This option is motivated by the fact that many powerful codes are inherently nonlocal~\cite{bravyi2010tradeoffs}, rely on random constructions~\cite{leverrier2015quantum}, or have structure in higher dimensions~\cite{scruby2024high}, in which case that structure is not necessarily retained when embedded in a 2D qubit lattice.

    Our generic placement algorithm begins by arranging the full graph in clusters or communities of nodes with short graph distances~\cite{blondel2008fast}. From this layout, we extract a heuristically maximal planar subgraph by iteratively adding edges with ascending length to a subgraph if the edges preserve planarity (Fig.~\ref{fig:algo}, step 2). This planar subgraph is then compactly embedded in 2D using a force-directed spring layout~\cite{kamada1989algorithm}, which minimizes the squared differences between the Euclidean and graph distances (Fig.~\ref{fig:algo}, step 3). While this does not guarantee planarity, we have empirically found the spring layout to result in predominantly planar layouts if a planar graph is provided as input. The layout is then rasterized and further compacted. Finally, the qubit-tier edges are routed as straight edges. The qubit-tier edges that cannot be routed crossing-free are saved for a higher tier. All remaining edges in the graph are also attempted in the qubit tier and are routed if they can be realized as crossing-free straight edges (Fig.~\ref{fig:algo}, step 4). 

    To route the remaining edges, a higher routing tier is created (Fig.~\ref{fig:algo}, step 5). Adding a routing tier corresponds to bump-bonding a chip onto the existing stackup, creating a new flip-chip interface (Fig.~\ref{fig:stackup}\textbf{b}). The edges are then iteratively routed as straight lines, where collisions are resolved using bump transitions (Fig.~\ref{fig:algo}, step 5). If a path becomes too congested for this routing approach, HAL employs the A* algorithm \cite{hart1968formal} to route around the obstruction while minimizing edge length. Collisions along this path are still resolved using bump transitions (see App.~\ref{App_algo} for details). Should the number of bump transitions within one edge exceed a user-defined maximum number, the edge is moved to a queue to be routed on a higher tier.

    If no more edges can be routed, the tier is considered congested. A new tier is created, and the remaining edges are reattempted. This approach is repeated until all edges have been successfully routed (Fig.~\ref{fig:algo}, dashed arrow).

    An alternative algorithm would partition the full connectivity graph into a minimal set of planar subgraphs and route the edges of each subgraph in its own crossing-free layer~\cite{tremblay2022constant}. Since qubit positions are fixed across all subgraphs, many edges may need to follow long polygonal paths in parallel to remain crossing-free \cite{pach1998embedding,schaefer2021new}. This may violate realistic hardware constraints, such as maximum coupler length and finite coupler width. We take advantage of bi-layer routing within one tier of a realistic multi-chip stackup and choose to lift the requirement of fully planar subgraphs. This allows edges to be much shorter, potentially at the cost of a few more layers. A detailed comparison between these two approaches is the topic of future work.

    \section{Layout examples}\label{sec:examples}

    We now turn to laying out several codes with HAL. In this section, we focus on three promising qLDPC code families: bivariate bicycle (BB) codes \cite{mackay2004sparse, kovalev2013quantum, bravyi2024high, yoder2025tour}, tile codes \cite{steffan2025tile, liang2025planar}, and radial codes \cite{scruby2024high}. In Sec.~\ref{sec:results} we also investigate directional codes~\cite{geher2025directional} and Tanner codes~\cite{leverrier2022quantum,radebold2025explicit}. This set covers both topological and non-topological code families.
    
    \textbf{Bivariate bicycle codes} are constructed from X- and Z-stabilizers that act on data qubits within a finite range. These stabilizers are translationally invariant and typically tiled across the surface of a torus. These codes have been proposed as leading candidates for fault-tolerant architectures, with a qubit overhead up to 10 times lower than in surface codes \cite{yoder2025tour}.
    
    The bicycle codes we focus on all feature a weight-4 nearest-neighbor lattice, with two additional long-range couplers per qubit, giving a total weight of 6. The long-range couplers result from either the structure of bulk stabilizers or the embedding of a torus on a planar surface, leading to periodic boundaries.

    \begin{figure*}[htbp]
      \centering
    
      \begin{minipage}{\textwidth}
        \centering
        \includegraphics[width=\textwidth]{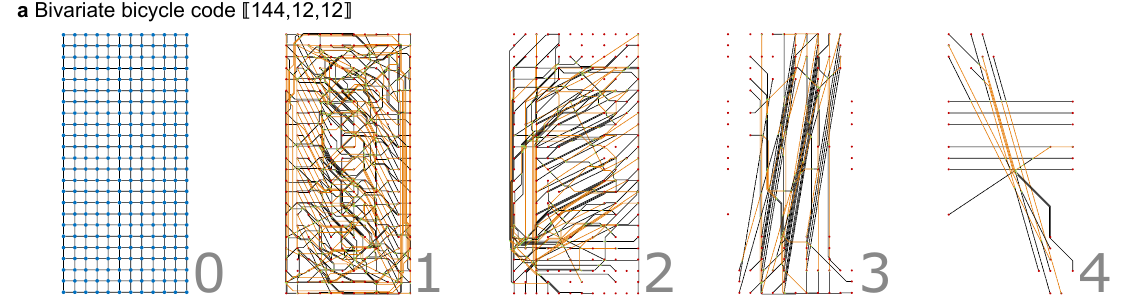}
      \end{minipage}
        \vspace{0.6em}
    
    \begin{minipage}{\textwidth}
        \centering
        \includegraphics[width=\textwidth]{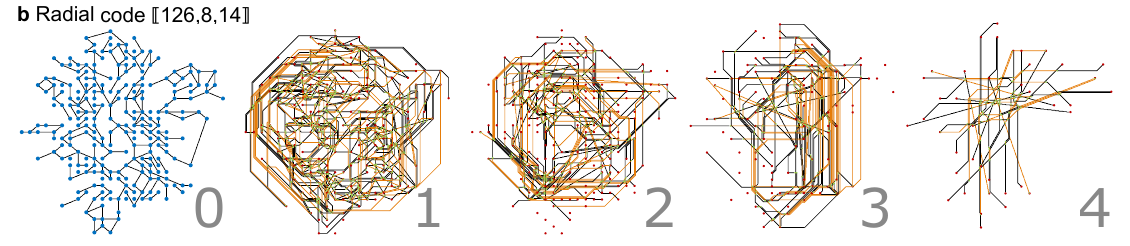}
      \end{minipage}
    \vspace{0.6em}

      \begin{minipage}{0.603\textwidth}
        \centering
        \includegraphics[width=\textwidth]{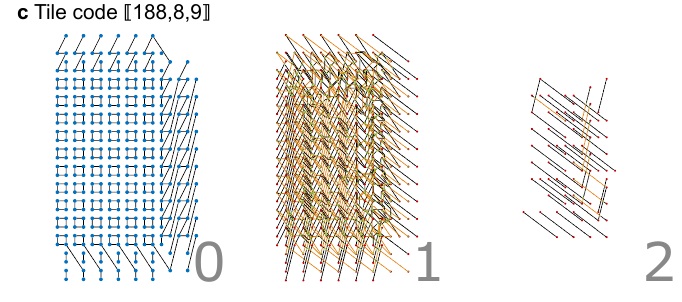}
      \end{minipage}
      \hfill
    \begin{minipage}{0.39\textwidth}
      \centering
      \renewcommand{\arraystretch}{1.4}
      \setlength{\tabcolsep}{7pt}
      \small
      \begin{tabular}{@{}l|cccc@{}}
        \toprule
        QECC & Tiers & Length & Bumps & TSVs \\
        \midrule
        BB  & 5  & 11.08 & 5.06 & 3.27 \\
        Radial & 5  & 13.19 & 5.30 & 3.16 \\
        Tile  & 3  & 2.98  & 2.89 & 2.17 \\
        \bottomrule 
      \end{tabular}
      \refstepcounter{table}
      \\[0.75em]
      \parbox[t]{\linewidth}{\raggedright \hspace{0.3em} TABLE~\thetable. Extracted hardware parameters.}
      \label{tab:routing_cost}
    \end{minipage}
    
      \caption{\textbf{Comparison of hardware layouts and hardware parameters across three QECCs.} The qubit positions of the bivariate bicycle (BB) code in panel \textbf{a} are chosen to realize a nearest-neighbor square lattice on the qubit tier. Higher tiers are depicted to the right, with the tier index indicated in the bottom right of each subfigure. The radial code in panel \textbf{b} does not have a topological structure. Therefore, we resort to HAL's default placement strategy of using a force-directed spring layout to maximize the number of short edges. Panel \textbf{c} shows the layout of a tile code, which has a similar structure to the BB code except that it is embedded on a plane with open boundaries rather than periodic ones. All codes require a weight-6 qubit connectivity and have comparable numbers of physical qubits. Table~\ref{tab:routing_cost} show individual hardware parameters for the three selected QECCs. While the BB and radial codes have similar values for all parameters, the latter have a slightly higher average edge length that can be explained by their lack of a nearest-neighbor qubit tier. On the other hand, the tile code shows significantly reduced values, most notably, about a fourfold reduction in the average edge length.}
    \label{fig:case_study}
    \end{figure*}

     We base our BB construction on the generalized toric construction introduced in Ref.~\cite{liang2025generalized}, which embeds bivariate bicycle codes on an optionally twisted torus. Here, one of the periodic boundaries implements a shift in the qubit rows, leading to novel and more efficient codes. Notably, Ref.~\cite{liang2025generalized} finds many twisted toric codes with significant boundary shifts, giving rise to tori with a high aspect ratio and, thus, narrow qubit lattices. We discuss the implications this has for optimal layouts in App.~\ref{App:exploit_struct}.

     In Fig.~\ref{fig:case_study}\textbf{a}, we lay out the \(\llbracket144,12,12\rrbracket\) gross code using HAL. We provide HAL with custom qubit positions to enforce the placement of qubits on a square lattice. This results in a weight-4 qubit tier with nearest-neighbor connectivity and four higher tiers to route the long-range couplers. The underlying tileable nature of the code, which arises from the translation-invariant local stabilizers, is visible in the many parallel edges. Furthermore, many couplers cross the entire lattice, a consequence of the periodic boundaries. These couplers cannot be routed on highly congested lower tiers. Instead, they are moved to more sparsely populated higher tiers.

     Notably, our layout requires more than two layers, despite the thickness-2 property shown in Ref.~\cite{bravyi2024high}. This is a consequence of allowing interlayer transitions along edges, which reduces edge length at the cost of more layers (see Sec.~\ref{sec:algo}). Our routing strategy also allows two-thirds of the edges to be realized in a weight-4 nearest-neighbor lattice, which is not the case if one partitions the connectivity graph into two planar subgraphs as in Ref.~\cite{bravyi2024high}.

     \textbf{Tile codes} rely on a general construction that implements BB codes on a planar surface with open boundaries, leading to true $\mathcal{O}(1)$-locality on a 2D planar lattice. Stabilizers are strictly defined within a bounded tile. When tiling a plane, the tiles that reach beyond the edge of the supported data qubit lattice are truncated. Data qubits and stabilizers are pruned appropriately to ensure commutativity and distance-preservation.

     The tile codes found in literature achieve lower overhead savings on average than BB codes. If compensated with a higher weight, e.g., a weight of 8, and high qubit counts, high-efficiency codes with order-of-magnitude qubit savings over the surface code can still be found.

     We lay out the \(\llbracket188,8,9\rrbracket\) tile code with HAL in Fig.~\ref{fig:case_study}c. Note that these codes do not always feature a weight-4 qubit tier with nearest-neighbor connectivity, as the tile code construction does not enforce this. Similarly, it does not specify the location of the check qubit within each tile. We devise different heuristics to choose a position and compare the resulting hardware complexities. In general, choosing a position where the sum of Euclidean distances between the check qubit and its supported data qubits are minimized performs consistently well. For more details, see App.~\ref{App:vary_check_qubit_pos}.

     The layout shows that the true $\mathcal{O}(1)$-locality and tileability allow for much shorter edges and even greater regularity than the gross code. From Tab.~\ref{tab:routing_cost}, we see that the average edge length is almost four times smaller than in the gross code. Most edges can be realized as straight edges, with a bump pattern that repeats throughout the lattice. The edge density is reduced toward the boundaries, as expected from the truncation of stabilizers along the boundary. The compactness of the routing also allows for roughly the same number of edges as the gross code to be routed in two fewer tiers.
     
     \textbf{Radial codes} are obtained from the lifted product of classical radial codes \cite{fossorier2004quasicyclic}. A classical radial code uses a pair of integers $(r,s)$ and can be visually arranged in $r$ concentric rings containing $s$ spokes. The quantum code is then formed from $r$ copies of a classical radial code responsible for the $X$-basis and $r$ copies responsible for the $Z$-basis, resulting in a quantum code with parameters \(\llbracket2r^2s,2(r-1)^2,\leq2s\rrbracket\). Each qubit is connected to $2r$ other qubits. While one may be able to identify geometric structure for quantum radial codes in 3D, it is unclear how much structure can be exploited when embedded in our proposed 2D multilayered architecture. We, thus, resort to our spring-layout placement algorithm.

    One key feature of radial codes is that they allow for single-shot or ``constant-depth'' decodability, which leads to time-overhead savings of up to a factor of $d$. This is especially significant as logical computation in qLDPC codes with many logical qubits is often serialized and can incur large time overheads over codes with a single logical qubit---an order of magnitude for gross codes \cite{yoder2025tour}.

     In Fig.~\ref{fig:case_study}\textbf{b}, we lay out the \(\llbracket126,8,14\rrbracket\) radial code with HAL. Despite lacking a regular sublattice, our spring-layout placement algorithm manages to compactly arrange qubits with many nearest-neighbor connections. While higher tiers similarly do not exhibit the same regularity as BB codes or tile codes, HAL manages to route all edges within a comparable number of tiers. The shape of these layouts is characteristic of the spring-layout algorithm, which returns similar-looking layouts even for instances of other code families.

    \section{Laying out many codes} \label{sec:results}

    To identify trends in the practicality of these code families, we rely on two metrics: the logical efficiency and the hardware complexity. The \emph{logical efficiency} \( \eta_L = k \cdot d^{2}\!/\!n \), is derived from the Bravyi-Poulin-Terhal bound \cite{bravyi2009no,bravyi2010tradeoffs}, which states that any 2D, geometrically local quantum code must satisfy 

    \begin{equation}
        kd^2=\mathcal{O}(n),
    \end{equation}

    \noindent where varying coefficients in $\mathcal{O}(n)$ can be used to compare the logical efficiency across different 2D codes. Since the rotated surface code has $\eta_L=1$ regardless of size, one can also interpret this quantity as an improvement factor over the efficiency of surface codes.

    Central to our work, however, is to evaluate codes based on their compatibility with multilayer superconducting hardware. We derive a \emph{hardware complexity} metric from the explicit layouts generated by HAL. For each layout, we extract four raw quantities and combine them to compute the hardware complexity: the number of tiers, the average edge length across higher tiers in units of the shortest edge length, the maximum average of bump bonds across all tiers, and the average number of TSVs per edge on higher tiers.
    
    Each quantity, \(q_i\), is linearly rescaled between a \emph{baseline} value, \(b_i\), (resulting in a score of 0) and an \emph{optimistic} value, \(p_i\), (resulting in a score of 1). Here, \(b_i\) reflect the state-of-the-art hardware architecture used to implement surface codes, while \(p_i\), reflects advanced fabrication capabilities we consider optimistically attainable in the near future. The individual hardware parameter, \(c_i\), is given by
    \begin{equation}
        c_i = \frac{q_i-b_i}{p_i-b_i}.
    \end{equation}
    
    \noindent The overall hardware complexity, \(C_\text{hw}\), is then given by computing the weighted arithmetic mean of all four individual hardware parameters, and adding it to one:
    \begin{equation}
        C_{\text{hw}} = 1 + \frac{\sum_{i} w_{i} c_{i}}{\sum_{i} w_{i}},
    \end{equation}

    \noindent so that \(C_{\text{hw}}=1\) denotes an ideal, planar, single-tier, nearest-neighbor layout, while higher values indicate increasing fabrication complexity. In all quoted values for $C_{\text{hw}}$, the baseline was set to 1 tier, unit average coupler length, and 0 bump transitions or TSVs; the optimistic values (which result in \(C_{\text{hw}}=2\)) were set to 5 tiers, 10 times longer long-range couplers than short-range couplers, 4 bump transitions, and 3 TSVs per coupler. We provide context and justification for these values in App.~\ref{App:Opt_values}. We use a uniform distribution of weights, \(w_i\), but vary the weights to study the contributions from each individual hardware parameter in App.~\ref{sec:cost_variation_robustness}

    \begin{figure*}
        \centering
        \includegraphics{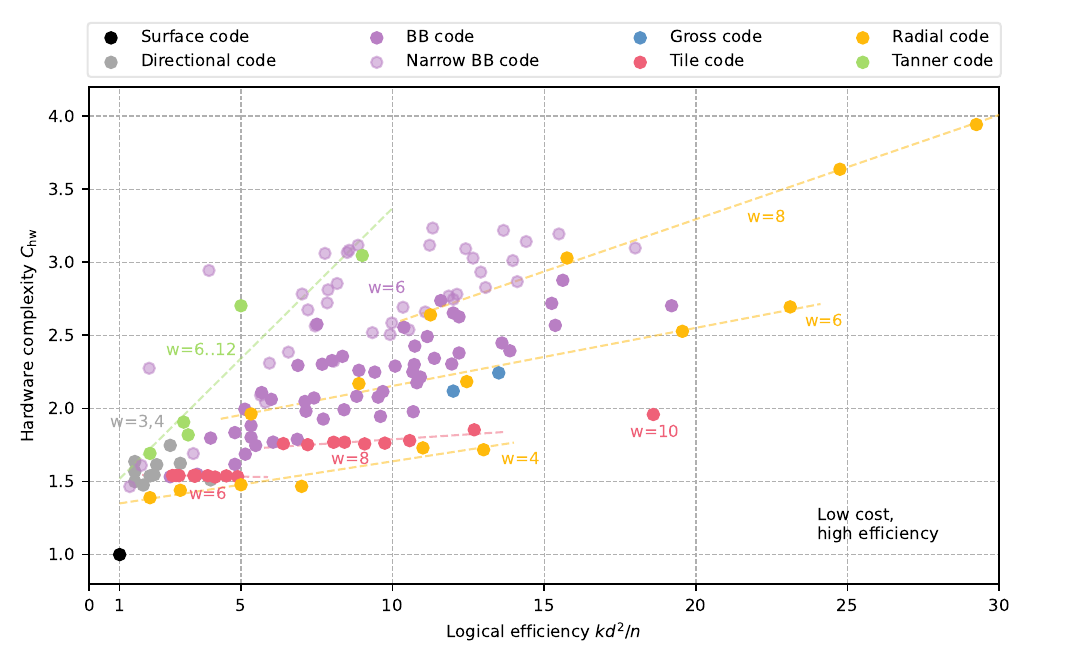}
        \caption{
        \textbf{Systematic comparison of directional, bivariate bicycle, tile, radial, and Tanner codes.} Points represent the logical efficiency and hardware complexity of a code instance. Dashed lines indicate linear fits for families of codes with equal check weight $w$, except for the Tanner codes, where the fit includes code instances of varying weight. Across the dataset, we observe a clear tradeoff between logical efficiency and hardware complexity, confirming that long-range coupling and complex connectivity are often essential to achieving high efficiency. Directional codes (gray) use iSWAP-gates to emulate long-range coupling; however, they require a toroidal embedding, leading to high hardware complexity at moderate efficiencies. BB codes (magenta) show higher hardware complexity scaling with size due to their periodic boundaries; narrow high-aspect-ratio BB codes (light magenta) benefit from spring layouts over square lattices. Tile codes (red) form distinct, flat bands in the plot, reflecting their modular, open-boundary layout; codes with higher weights achieve better efficiencies while incurring higher hardware complexities. Radial codes (yellow) match BB codes, with some weight-4 instances potentially outperforming all others in efficiency at minimal hardware complexity. Tanner codes (green) have the highest hardware complexity for a given logical efficiency, though they also match some BB codes. All code layouts are summarized in Tab.~\ref{tab:full_database} and can be individually inspected using our online database \cite{hal_database}.
        }
        \label{fig:complexity}
    \end{figure*}

    We use HAL to generate explicit layouts for roughly 150 code instances spanning the families mentioned in Sec.~\ref{sec:examples}. All code layouts can be individually inspected using our online database \cite{hal_database}. 
    
    Across the entire dataset, there consistently seems to be a tradeoff between logical efficiency and hardware complexity. We also observe many codes to have a hardware complexity around or below 2, which corresponds to saturating the optimistic fabrication capabilities. We anticipate these codes to be realizable in hardware in the near term. Further, there seems to be a non-negligible offset in the hardware complexity in the layout of even the simplest code compared to the surface code. This is due to the fact that all these codes rely on some physical long-range couplers, requiring at least one additional tier and an increased coupler length.
    
    \textbf{Directional codes} rely on iSWAP gates during syndrome-extraction circuits to circumvent the need for physical long-range coupling to measure long-range stabilizers \cite{geher2025directional}. As a result, these codes can be realized on a square or even hex lattice. However, the lattice is embedded on a torus, leading to periodic boundaries when realized in multilayered hardware. As expected, the boundaries cause these codes (shown in gray) to have a high hardware complexity compared to the moderate gains in logical efficiency. 

    \textbf{Bivariate bicycle codes.} We observe that the \(\llbracket144,12,12\rrbracket\) gross and \(\llbracket288,12,18\rrbracket\) two-gross codes (shown in blue) are among the best performing BB codes (magenta). This high performance is likely owed to the fact that they achieve a high efficiency with a low number of total qubits, and, thus, a low number of total edges. 

    We laid out all BB codes using a square grid layout (dark magenta) and the spring-layout algorithm (light magenta), as shown in Fig.~\ref{fig:complexity}, choosing whichever strategy gives the better hardware complexity on a case-by-case basis. In App.~\ref{App:exploit_struct}, we find the spring layout to outperform the square layout for codes with a high aspect ratio. Codes for which this is the case are featured in Fig.~\ref{fig:complexity} as ``narrow" codes, with aspect ratios of around 8 and higher. Note that the best BB codes are still those where the square layout has a lower hardware complexity.

    \textbf{Tile codes} seem to strictly outperform BB codes in hardware complexity for the same logical efficiency. Furthermore, we observe separate, flat bands forming. The code instances within each band are derived from the same underlying tile shape but tiled across an increasingly larger surface, leading to increasing logical efficiency. The flatness of these bands agrees with the intuition that the open boundaries and tileable nature of tile codes allow the hardware complexity to remain roughly constant regardless of size. On the other hand, BB codes have periodic boundaries, leading to couplers that increase with the size of the code, possibly explaining the higher slope in hardware complexity among BB codes compared to tile codes.

    Note that the different tile code bands correspond to different tile patterns, particularly different check weights. The weights, ranging from 6 to 10, seem to contribute directly to an increasing logical efficiency while incurring a constant offset in the hardware complexity. A higher weight leads to increased tier congestion and the need for more tiers in total. However, from inspecting the layouts of even the weight-10 code, we see that coupler regularity is well preserved across all tiers. 

    Generally, tile codes with high logical efficiency seem rare and are mostly obtained with high weights and qubit counts. Consequently, it would be interesting to apply weight-reduction techniques \cite{mcewen2023relaxing, shaw2025lowering, geher2025directional, zhou2025louvre,zhao2025simple} to tile codes, as well as physically demonstrate high-weight qubits.

    \textbf{Radial codes}. Similarly to the tile codes, we observe distinct bands forming (shown in yellow), corresponding to an increasing check weight. Each higher band exhibits a discrete increase in hardware complexity and a higher slope. Many radial codes achieve a hardware complexity that is comparable to that of BB codes. This observation broadly implies that the hardware requirements of BB codes, or the gross codes, is compatible with the needs of radial codes, despite their lack of topological structure.
    
    Interestingly, we find that the lowest band achieves high logical efficiencies with hardware complexities that outperform all other codes. This achievement is likely due to a low weight of four and a small qubit count. A low check weight would bear further advantages. Each edge represents a physical coupler that can fail, needs to be controlled individually, and constrains the optimal calibration of a large processor \cite{klimov2024optimizing}. Further, a shorter weight allows syndrome extraction to have low depth. Combined with the constant-depth decodability of radial codes, these codes could have higher logical clock rates than BB codes or tile codes. Finally, codes with lower stabilizer weights may be more accessible to certain promising qubit types, such as fluxonium qubits, which have less available capacitance for coupling to neighboring qubits than transmon qubits~\cite{krantz2019quantum, ding2023high}.

    Note, that we were only able to find radial codes achieving a maximum distance of $2s$ for a subset of the $(r,s)$-pairs we searched. However, there is reason to believe that $d=2s$ radial codes exist for all $(r,s)$-pairs. In App.~\ref{app:dist_est_radial}, we describe our search for radial codes in detail and provide justification for using the hardware complexity of a lower-distance radial code as a proxy for a high-distance code with the same $(r,s)$-values. 

    Our investigation stresses the importance of finding high-distance, low-weight radial codes. This could imply searching for more code instances or investigating why the lifted product construction, which is used to generate radial codes, sometimes introduces low-weight logical operators and how to prevent this from happening~\cite{scruby2024high}.

    \textbf{Tanner codes} are a family of asymptotically good qLDPC codes, achieving a constant encoding rate and a distance linear in the number of physical qubits \cite{leverrier2022quantum, guemard2025moderate, radebold2025explicit}. It is worthwhile to investigate the performance of smaller explicit instances. Using our generic spring-layout approach, we lay out promising instances of Tanner codes found in Ref.~\cite{radebold2025explicit}. Their hardware complexity (shown in green) lies at the upper end of the spectrum of codes for a given logical efficiency, though they are comparable to some BB codes. 
    
    This is not entirely surprising, as the connectivity graphs of Tanner codes have expansion properties, making them highly nonlocal. These codes also have varying stabilizer weights,  ranging from 6 to 12, even within one code instance. Reducing these weights using general techniques \cite{zhao2025simple} could alleviate this overhead. Finding experimentally accessible, small instances of Tanner codes represents a valuable effort as they also provide other potential advantages, such as single-shot decodability~\cite{gu2024single}.
    
    \section{Outlook}

    In this work, we present HAL, a heuristic, run-time-efficient tool that automates and optimizes the placement and routing of arbitrary QECCs on superconducting hardware. HAL assumes a multilayer stackup with long-range coupling. Following a sequence of heuristic algorithms, HAL extracts a planar subgraph from a QECC connectivity graph, places the nodes of this subgraph onto the qubit tier using a spring layout, and rasterizes them. It routes the edges of the planar subgraph on the qubit tier and then proceeds to route the remaining edges on higher tiers using a modified A* algorithm. This algorithm incorporates the freedom for edges to transition between layers within one tier using bump bonds, and moves edges to higher tiers using TSVs once a tier becomes congested.

    We use HAL to study the hardware complexity of various qLDPC code families, generating explicit layouts for nearly 150 codes across several qLDPC code families. We find BB codes to benefit somewhat from their regular structure, but to suffer from the periodic boundaries. Tile codes overcome this problem, achieving true locality and regularity, but suffer from reductions in logical efficiencies. This needs to be compensated with a high qubit number and high weight, prompting the investigation of weight-reduction techniques. We find many radial codes that are competitive with BB codes in terms of both logical efficiency and hardware complexity. We also find low-weight instances of radial codes to be hardware-friendly and potentially very efficient, motivating further study of their construction.

    Follow-up work may include improvements to the HAL algorithm or its adaptation to different hardware constraints and qubit modalities. We are also interested in studying the increase in hardware complexity when augmenting a qLDPC code with an ancilla graph to enhance its computational capabilities. Another promising application of HAL is the proactive discovery of QECC architectures optimized for hardware feasibility by design. Thanks to its automated nature, HAL’s hardware complexity estimates could be incorporated into the cost function of a reinforcement learning agent tasked with code discovery.

    Our work also highlights the importance of advancing superconducting qubit fabrication. While we assume a stackup informed by current superconducting qubit technology, there remains significant potential in developing novel 3D integration techniques, such as compact, dedicated multilayer routing modules for coherent long-range coupling. It is also valuable to pursue demonstrations of high-weight parity checks, which are especially challenging for certain qubit types such as fluxonium. Advancing these fabrication capabilities in parallel with the discovery of hardware-efficient QECCs will be essential to sustaining the competitiveness of superconducting qubits and driving the field toward low-overhead, fault-tolerant quantum architectures.

    \section{Author contributions}

    MM developed the computational framework and led the design and execution of core simulations that underpin the results. LP and DP originated the idea, provided conceptual guidance throughout the project, and contributed equally to generating the final results. VLA and CT provided essential support in developing the framework and executing the simulations. WDO and JAG supervised the project. MM, DP, and LP wrote the manuscript with input from all authors.
    
    \section{Acknowledgments}
    We gratefully acknowledge Chris McNally and Max Hays for fruitful discussions and Stergios Koutsioumpas and Joschka Roffe for carefully reading the manuscript.
    The authors acknowledge the MIT Office of Research Computing and Data for providing high-performance resources that have contributed to the research results reported within this paper.
    This material is based upon work supported by, or in part by, the U.S. Army Research Laboratory and the U.S. Army Research Office under contract/grant number W911NF2310255, and in part by the Intelligence Advanced Research Projects Activity (IARPA) and the Army Research Office, under the Entangled Logical Qubits program, and was accomplished under Cooperative Agreement Number W911NF-23-2-0212.
    VLA acknowledges support from the U.S. Department of Energy, Office of Science, Office of Advanced Scientific Computing Research, Department of Energy Computational Science Graduate Fellowship under Award Number DE-SC0025528.
    The views and conclusions contained in this document are those of the authors and should not be interpreted as representing the official policies, either expressed or implied, of IARPA, the Army Research Office, or the U.S. Government.

    \section{Code Availability}

    The code used to run and produce layouts of QECCs is available at \cite{hal_github}. A database of all laid out codes is provided in Table~\ref{tab:metrics_database} and can also be viewed at \cite{hal_database}.

    \appendix
    
    \section{HAL Algorithm}\label{App_algo}

    In this section, we describe the algorithm in closer detail, beginning with the placement of the nodes.
    
    \subsection{Placement phase}

    In the placement phase, we embed the code connectivity graph \(G=(V, E)\), where V is the set of nodes and E is the set of edges, on a regular lattice such that (i)~all nodes occupy distinct grid points and (ii)~a heuristically maximal subset of edges can later be routed in the plane without crossings. The procedure comprises three consecutive steps: heuristic maximal planar subgraph extraction, systematic integer realization of the planar subgraph, and rasterization and grid normalization.

    For codes that exhibit strong inherent structure---e.g., 2D topological codes~\cite{aharonov2011complexity, bravyi2003commutative, kitaev2003fault}, such as bivariate bicycle codes~\cite{liang2025generalized}---the user may supply custom node positions to enforce that structure.  When such positions are provided, only the rasterization and grid normalization step is performed during the placement phase.
    
    \subsubsection{Heuristic maximal planar subgraph extraction} 
    HAL begins the placement phase by extracting a large planar subgraph of the connectivity graph that can be routed entirely on the qubit tier. Although finding a true maximum planar subgraph is NP-hard, an incremental heuristic delivers solutions that are empirically within a few percent of the optimum while running in linear time~\cite{chimani2016note}.
    
     We apply the greedy Louvain community detection algorithm~\cite{blondel2008fast}, which arranges the graph in a 2D layout of locally connected communities, i.e., clusters of nodes with short graph distances. For each edge, we then extract whether it is an intra- or inter-community edge and the Euclidean length of the straight segment between its endpoints.  Edges are sorted: first, all intra-community edges are ordered by increasing length, followed by the inter-community edges, again from short to long.  Short intra-community edges tend to lie entirely inside local clusters and are unlikely to cross.  Long, inter-module links are the most likely to create crossings and can be routed in higher tiers if necessary.

    The algorithm scans this ordered list of edges once. Starting from an empty graph, it tentatively inserts the next edge and performs a Hopcroft–Tarjan planarity test~\cite{hopcroft1974efficient}; if the edge preserves planarity, it becomes part of the subgraph, otherwise it is discarded and stored for higher-tier routing.  
    
    \subsubsection{Rasterized spring layout of the planar subgraph} 
    The heuristic maximal-planar subgraph extraction produces a planar subgraph \(G_{0}=(V,E_{0})\), where $E_{0}$ is a set of planar edges. We embed this subgraph in 2D with a Kamada–Kawai spring layout~\cite{kamada1989algorithm}. The Kamada–Kawai algorithm minimizes a global energy function that penalizes the squared differences between graph-theoretic distances and their corresponding Euclidean distances in the layout. The result is a drawing with (i)~nearly uniform edge lengths, (ii)~well-balanced angles, and (iii)~very little area wasted inside modules, all of which translate into shorter wires and fewer conflicts during routing~\cite{kamada1989algorithm}. 
    
    Next, the nodes are rasterized to integer lattice points while preventing double occupation.  A two–phase, purely combinatorial procedure achieves this goal while keeping the displacement of each node minimal.

    \emph{Phase 1 – naïve rounding and immediate acceptance.}
    Every node is mapped to its nearest lattice point \(\tilde p_v=(\lfloor x_v\rceil,\,\lfloor y_v\rceil)\). If \(\tilde p_v\) is not claimed by any other node, the placement is accepted and the site is marked \emph{occupied}.

    \emph{Phase 2 – priority conflict resolution.} Nodes still in conflict enter a min-heap keyed by the Euclidean distance to the nearest currently free lattice site. The heap is processed greedily: the node that can stay \emph{closest} to its preferred position is removed, the nearest free site is found by expanding square shells of increasing radius, and the node is fixed there.  Whenever a site is occupied, the distance keys of the remaining heap elements are updated in place.
    
    \subsubsection{Compaction and grid normalization}
    After the nodes are placed, empty rows or columns may remain in the lattice. A final rasterization pass removes this slack. The set of distinct \(x\)-coordinates is sorted, and the \(i\)-th element is mapped to \(i\); the same is done for the \(y\)-coordinates. The monotone remap preserves the embedding planarity and relative edge lengths measured in grid units while compressing the footprint. The resulting coordinates are translated to the positive quadrant and scaled to a user-defined device size as a final step. 

    \subsection{Routing phase}

  The routing phase assigns an explicit geometric path to every edge in the connectivity graph. It proceeds tier by tier, starting with the qubit tier and creating further tiers on demand up to the user-specified maximum number of tiers.
   
    Although the heuristic maximal-planar-subgraph (MPS) step guarantees that all edges placed on the qubit tier can, in principle, be drawn as straight segments without crossings, a spring layout of the MPS does not guarantee this. We observe empirically that our spring layout algorithm tends to produce predominantly planar layouts when given planar graphs as input. Nodes already occupy distinct cells of a 2D grid; these cells are marked \textit{blocked} and thus unusable.  The router scans the MPS edges once and paints the corresponding grid cells along the straight line between their endpoints. Whenever this succeeds, the edge is declared routed; otherwise, it is enqueued in a FIFO (First-In-First-Out) that initially contains all edges \emph{not} in the MPS. After this, all other edges are also attempted as straight lines on the qubit tier, even if they were not part of the MPS. This procedure allows us to maximize the number of edges placed on the qubit tier.
    
    After laying out the qubit tier, any edge still in the FIFO requires a higher routing tier to be routed. All nodes incident to such edges are copied to a fresh \mbox{$(x,y,z)$} grid that represents the first routing tier.  In hardware, this is realized with a TSV providing a vertical connection from the qubit on the qubit tier to a higher routing tier, which has been bump-bonded onto the bottom chip.
    
    Every routing tier is modeled as a three-dimensional occupancy grid \(\mathcal{G}\subset\mathbb{Z}^{2}\times\{0,1\}\) whose slices \(z=0\) and \(z=1\) represent the bottom and top layers in a flip-chip geometry (Fig.~\ref{fig:stackup}\textbf{a}). A vertical transition between the slices is realized with a bump bond. Cells that correspond to nodes or already-committed traces are blocked. A user-defined expansion value expands every newly accepted trace by an additional safety margin so that subsequent routes maintain the required spacing.
    
    Within each routing tier, edges are processed in a fixed order that is not necessarily globally optimal but has proven highly robust: straight-line edge length. Edges are sorted in \emph{ascending} order of this estimate, thereby routing ``easy'' edges first. 
    
    HAL first attempts to connect the endpoints for each edge along a straight line. If the path is obstructed but the opposing layer is free, the edge performs a bump-bond transition to the opposing layer and only switches layers again at the subsequent obstruction or if the target node is reached.  This approach inserts the minimum number of bump bonds compatible with a straight line route, which minimizes edge length.
    
    If this routing attempt fails, the edge is retried with an A* search~\cite{hart1968formal}, which operates on the same grid but explores a larger neighborhood.  Potential successor moves are the four adjacent cardinal displacements, the four adjacent diagonal displacements within the current layer, and one vertical transition. The heuristic is the Euclidean distance to the target; it remains admissible and consistent in the presence of vertical moves. A successor is discarded when (i) it would enter a halo, i.e., a restricted area, or (ii) it lands on a cell already used by a trace on the same layer.
    
    Routing an edge can fail in two ways: (i) neither algorithm finds a path, or (ii) a found path violates the important hardware rule that no route may contain more than a user-defined maximum number of bump transitions. Physically, each bump transition may lower the coupler quality factor, reducing its coherence. Motivated by recent experiments, we set the maximal number of bump transitions per edge to 10 in all datasets reported here (see App.~\ref{App:Opt_values} for more details). A failed edge is appended to the back of the current FIFO. When the router pops an edge already attempted in the present tier, it declares the tier congested, freezes its traces, creates a new empty grid, copies the still-unrouted nodes to that grid, and starts the process anew with the carried-over FIFO. Routing terminates once the FIFO is empty, meaning every edge has an assigned path that respects all geometric and technological constraints.

    \subsection{User-configurable settings} \label{user_configurable_settings}
    HAL exposes several user-configurable parameters to tailor the placement and routing process to specific hardware constraints:
    \begin{itemize}
        \item \textbf{Custom positions:} an explicit map $p_0:V\!\rightarrow\!\mathbb{Z}^{2}$ that overrides the automatic placement for vertices.
        \item \textbf{Edge margin:} The safety margin (in grid cells) around every routed trace. We use a default value of 1.
        \item \textbf{Node size:} The radius added around each node before routing starts; reserves area for local wiring, which interconnects must not overrun. We use a default value of 1.
        \item \textbf{Grid size:} Determines the overall device area, aspect ratio, and the granularity of the layout canvas. We use a default value of 500.
        \item \textbf{Maximum bump transitions per coupler:} Limits the number of bump transitions for any single connection. When violated, the edge is popped to the next tier. We use a default value of 10.
        \item \textbf{Maximum TSVs per coupler:} Restricts the number of through-silicon vias per connection. When violated, the routing fails and aborts. We do not use a default value and ignore this restriction in the results we present in this paper as it would make some codes impossible to layout. 
        \item \textbf{Maximum coupler length:} Limits the maximum connection length between two qubits/nodes. When violated, the routing is popped to the next tier. We use a default value of 1000 times the smallest coupler length. 
    \end{itemize}

    \section{Runtime Analysis}

    \begin{figure*}[t]
        \centering
        \includegraphics{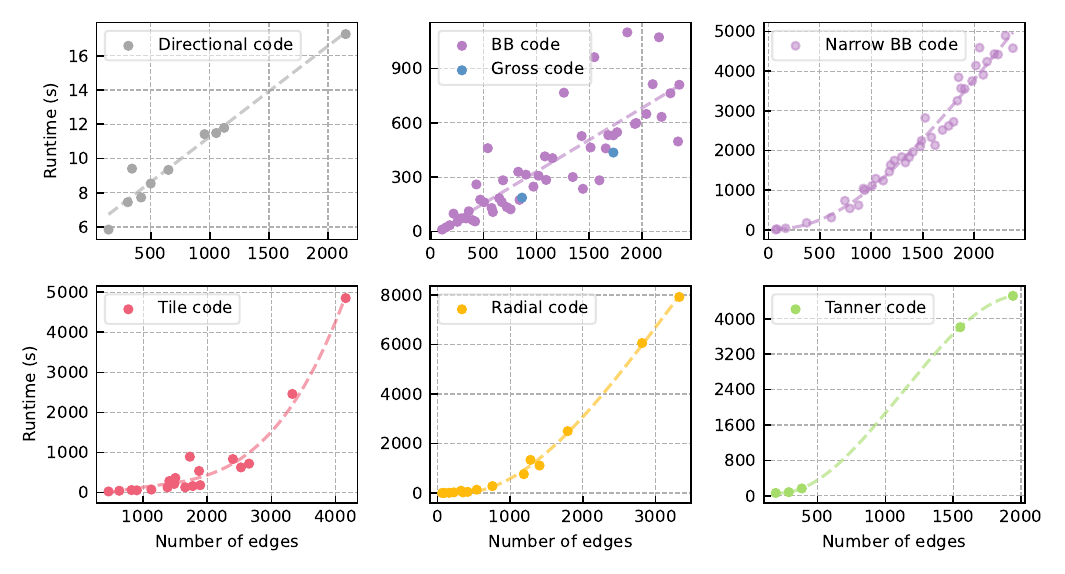}
        \caption{\textbf{Runtime time of HAL as a function of the number of edges in the input connectivity graph.} Dashed lines represent polynomial fits to the data. A linear fit was performed on the directional and BB codes, and degree-3 polynomial fits otherwise. The good fit results suggest that HAL has an efficient, polynomial runtime.}
        \label{fig:time_complexity}
    \end{figure*}

    To characterize the computational complexity of HAL, we performed a scaling analysis to understand how the runtime behaves as a function of the input problem size. We define the problem size by the number of edges in the connectivity graph of the quantum error correcting code being processed. 

    We measured the total time required for the placement and routing of the codes from Fig.~\ref{fig:complexity}. The execution time was measured in seconds, and the tool was run using only a single core of a 13th Gen Intel Core i7 CPU with 64GB of RAM. The results of this analysis are presented in Fig.~\ref{fig:time_complexity}. Note that runtimes increase with a higher grid size. We use a grid size of 500.

    As depicted in Figure \ref{fig:time_complexity}, we observe a clear trend in the relationship between the execution time and the number of edges. We perform a linear fit to the runtimes of the directional and BB codes, and degree-3 polynomial fits to the other datasets, achieving an R2 goodness score of more than 0.95 for all codes except for the BB codes, where some outliers lowered the R2 score to around 0.75. Note that when fitting exponential functions to the data, we obtain significantly worse fit performance, with most codes achieving R2 scores of around 0.7. In this case, the narrow BB codes and radial codes even obtain negative values for R2, indicating that the value of each sample is better predicted using the mean of the data than the fit.
    
    This numerical evidence suggests that HAL has a polynomial runtime in the relevant input size and the number of edges. Due to the heuristic nature of HAL, we expect this scaling to continue for larger codes as well. Furthermore, the most extended runtime of an individual code, the \(\llbracket416,18,26\rrbracket\) radial code, was only 2 hours and 13 minutes. This code is already significantly larger than QECCs typically studied, e.g. in Ref.~\cite{bravyi2024high}, underscoring that HAL has fast runtimes for typical code sizes. 
  
    \section{Rescaling Extracted Hardware Parameters}\label{App:Opt_values}

    Central to our work is to evaluate codes with respect to their compatibility with our proposed multilayered architecture. We refer the reader to Sec.~\ref{sec:results} for our definition of the hardware complexity. Although we do not explicitly include qubit count or check weight in our hardware complexity model, their influence shows up in the resulting complexity of the layouts HAL produces. When computing the total hardware complexity, we consider only parameters directly determined by the layouts---this also prevents double-counting causes as effects.

    When calculating the overall hardware complexity, we must choose the appropriate baseline and optimistic values to rescale the extracted hardware parameters. Baseline values are set to the hardware requirements of surface code architectures, implying one tier, a normalized coupler length of 1, and 0 bump bonds and TSVs per coupler. Optimistic values are chosen according to recent demonstrations of novel hardware capabilities.
    
    In this appendix, we cite the manuscripts we use to derive our optimistic values. In App.~\ref{sec:cost_variation_robustness} we provide further justification for our choices by demonstrating that our overall conclusions are robust to $\pm50~\%$ variations in the optimistic values. Table \ref{tab:opt_values} lists the optimistic values we chose for the hardware parameters and the relevant citations.

    \textbf{Tiers}. Ref.~\cite{rosenberg20173d} presents proof-of-principle experiments for an architecture of 3 vertically stacked chips using bump bonds, forming several routing interfaces. In the proposed architecture, the routing tiers are used to route control and readout circuitry and to place the qubits. Using similar bump-bonding technology, we anticipate that stacking 4 more chips on top of this stackup will be possible in the near future. This stackup would realize 5 routing tiers for long-range couplers, including the qubit tier. We set the optimistic value for the number of tiers to 5.

    \begin{table*}[t]  
        \small
        \renewcommand{\arraystretch}{1.3}
        \begin{tabular}{
            >{\raggedright\arraybackslash}p{5cm}
            >{\centering\arraybackslash}p{3.6cm}
            >{\raggedright\arraybackslash}p{5.5cm}
        }
            \toprule
            Parameter & Optimistic value & Representative citation \\
            \midrule
            Number of tiers    & 5  & \cite{rosenberg20173d}: 3 chips stacked to route control and readout signals and place qubits\\
            \makecell[l]{Coupler length \\ (in units of short-range coupler)} & 10 & \cite{wang2025demonstration}: 6.5~mm-long coupler demonstrated with 99.2~\% fidelity \\
            
            Bump transition per coupler & 4 & \cite{field2024modular}: Coupler interrupted by 4 bumps demonstrated with 99.1~\% fidelity \\
            TSV per coupler     & 3 & \cite{ hazard2023characterization}: Qubit embedded in TSV with a quality factor of $750\times10^3$ \\
            \bottomrule
        \end{tabular}
        \caption{\textbf{Optimistic values used for benchmarking hardware complexity}. These values reflect ambitious but plausible near-to-mid-term hardware capabilities. A layout that requires the optimistic value for all parameters achieves a hardware complexity of 2.}
        \label{tab:opt_values}
    \end{table*}

    \textbf{Length.} Ref.~\cite{wang2025demonstration} demonstrated a small BB code with couplers that are between 1~mm for nearest neighbor connections and up to 6.5~mm for long-range couplers with average two-qubit gate fidelities of 99.2~\%. In \cite{xiong2025scalable}, a two-qubit gate mediated by a 1.14~cm long coupler was realized with a fidelity of 99.37~\%. With nearest-neighbor connections that are 0.5-\SI{1}{\milli\meter} long, these two works along with other works \cite{heya2025randomized, xu2025tunable} provide strong indication that high-fidelity on-chip long-range couplers that are ten times longer than nearest-neighbor connections will be realizable. We set our optimistic value accordingly.

    \textbf{Bump bonds.} Ref.~\cite{field2024modular} realized a two-qubit gate using a coupler interrupted by 4 bump transitions with a fidelity of 99.1~\%. The authors did not observe a significant reduction in fidelity compared to two-qubit couplers with fewer bump interruptions. Bump bonds can always incur a reduction of the internal quality factor of the coupler, and will eventually limit the lifetime and, thus, fidelity of the two-qubit gate. For the case of 4 bump bonds, these quality factors seemed to have been high enough to be negligible. We set our optimistic value for bump bond interruptions to 4.

    \textbf{TSVs}. To our knowledge, a TSV-interrupted two-qubit coupler has not been reported in the literature at the time of writing this manuscript. Ref.~\cite{hazard2023characterization} realized a design where the qubit derives most of its capacitance from a TSV, by being effectively embedded in a TSV. This qubit design showed an average quality factor of $750\times10^3$. 
    
    We can use this number to estimate the limit on the two-qubit gate fidelity imposed by TSV interruptions. We begin by calculating the coupler lifetime $T_{1,\text{cplr}}$:

    \begin{equation}
    \begin{split}
        Q_{\text{cplr}}&=\frac{n}{Q_{\text{TSV}}}\\
        T_{1,\text{cplr}}&=\frac{Q_\text{cplr}}{\omega_\text{cplr}}
    \end{split}
    \end{equation}

    \noindent where $Q_{\text{cplr}}$ is the quality factor of a coupler that is interrupted by $n$ TSVs, and $\omega_{\text{cplr}}$ is the frequency of the coupler. Here, we assume that the coupler lifetime is only limited by the presence of TSVs.

    The extent to which the coupler lifetime limits the two-qubit gate fidelity depends on the specific two-qubit gate scheme. We provide a worst-case estimate of the contribution of the coupler lifetime to the fidelity of the two-qubit gate, assuming that it is a direct limiting factor \cite{ding2023high}:

    \begin{equation}
        F_{2\text{qb}}=1-\frac{4t_g}{5}\frac{1}{T_{1,\text{cplr}}}
    \end{equation}

     \noindent where $t_q$ is the gate duration. Assuming $t_g = \SI{70}{\nano\second}$, $\omega_\text{cplr}/2\pi = \SI{7}{\giga\hertz}$, $n = 3$, and a TSV quality factor $Q_{\text{TSV}}=750\times10^3$, we find $T_{1,\text{cplr}} = \SI{5.7}{\micro\second}$ and $F_{2\text{qb}} \approx 99\%$. For realistic gate schemes, we expect the coupler lifetime to have a much weaker impact on two-qubit gate fidelities, particularly if the coupler mediates only a virtual exchange coupling. TSV quality factors are also expected to continue improving. We therefore assume an optimistic scenario of 3 possible TSVs per coupler.

    Note that the hardware complexity is defined such that $C_\text{hw}=1$ for all surface codes regardless of their size. This relation emerges because $C_\text{hw}$ is calculated using averaged contributions from coupler length, bump bonds, and TSVs. The motivation for this is to mirror the size independence of the logical efficiency, which is constant at 1 for all sizes of the rotated surface code. The hardware complexity can also be understood as an evaluation of a code's tileability. Since a surface code is perfectly tileable and the complexity of each tile does not change with the size of the code, its hardware complexity stays constant at $C_\text{hw}=1$. Similarly, since tile codes are very well tileable, their hardware complexity remains constant even when tiled across a larger area (see Fig.~\ref{fig:complexity}).

    \section{Robustness to Variation\\in Hardware Complexity Model}\label{sec:cost_variation_robustness}
    
    \begin{figure*}
        \centering
        \includegraphics{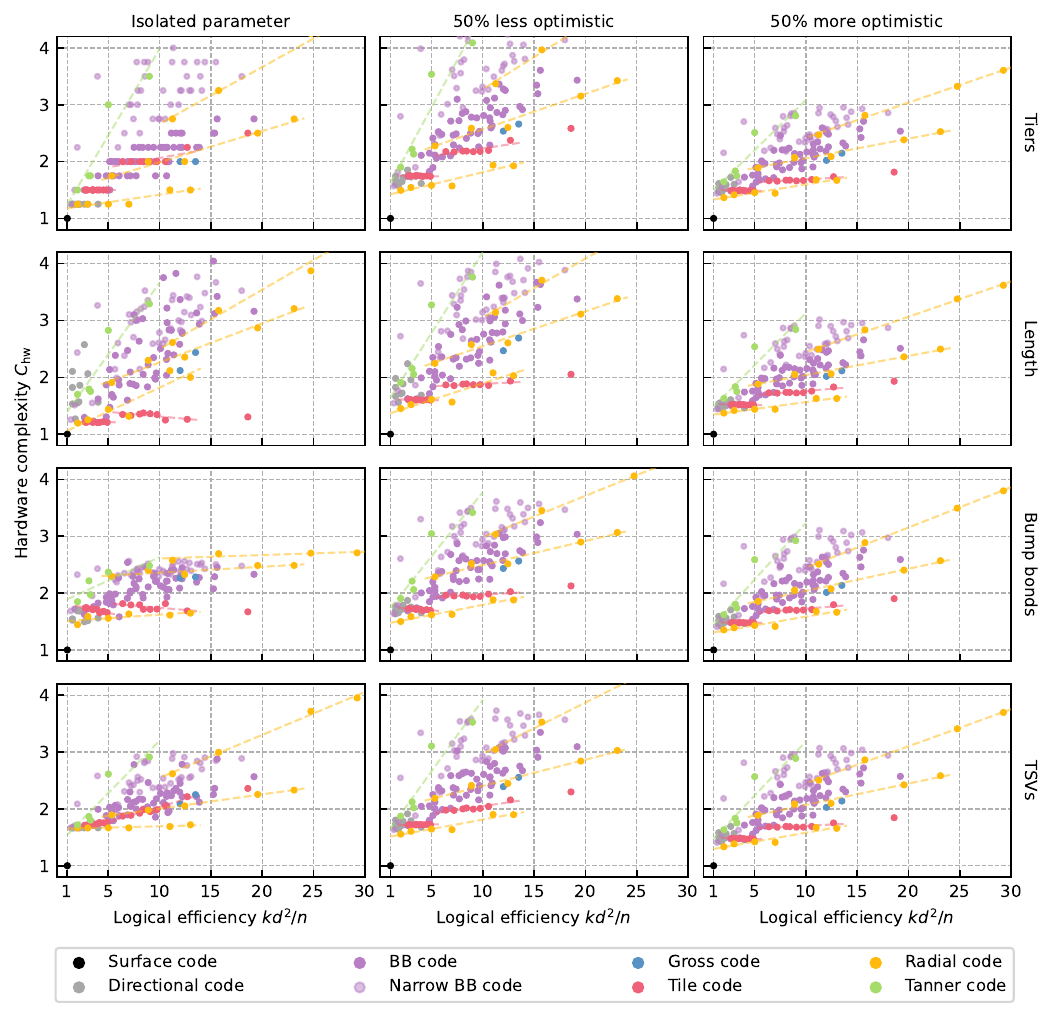}
        \caption{\textbf{Dependence of results on variation in hardware complexity model.} Each row studies a different individual hardware parameter. In the leftmost column of each row, we set the weight of the corresponding parameter to 1, and all others to 0, to study the contributions of the parameter in isolation. This study reveals that each of the four parameters highlights a different component of the hardware complexity, and the combination of all four is required for a holistic evaluation of a code. In the second and third columns, the weights are reset to a uniform distribution, and the optimistic value for the given parameter is varied by \(\pm50\%\). The overall trends from the results shown in Fig.~\ref{fig:complexity} are preserved with minor relative changes, highlighting the robustness of the hardware complexity framework in HAL to uncertainty in parameter values.}
        \label{fig:cost_variation_robustness}
    \end{figure*}    

    The model we use to calculate the hardware complexity underpins the conclusions we draw from the final results in Sec.~\ref{sec:results}. Therefore, it is crucial to study how sensitive the results are to changes in the complexity model. 

    In Fig.~\ref{fig:cost_variation_robustness}, we vary the weights, \(w_i\), and optimistic values, \(p_i\), that enter the hardware complexity (see App.~\ref{App:Opt_values}) to study the robustness of our claims to uncertainty in these parameters, as well as break down the contributions from each complexity, \(c_i\), to the overall hardware complexity, \(C_\text{hw}\).

    In the leftmost column, we isolate each parameter, \(c_i\), in the row corresponding to the annotation on the right side of the figure by setting its weight to 1 and all other weights to 0. 
    
    \textbf{Tiers.} When isolating the contributions from the number of tiers, a discrete structure emerges, which agrees with the number of tiers being an integer metric. The overall trends from the main result in Fig.~\ref{fig:complexity} are still visible: the tile and radial codes are grouped in bands ordered by the degree of the nodes. Directional and weight-4 radial codes have the fewest tiers, followed by tile codes, BB codes, and Tanner codes.
    
    \textbf{Length.} In the next row, the main strength of the tile codes becomes evident: the average edge length is short and does not increase significantly with improving logical efficiency. This property is made evident by the increasing gap between the hardware complexities of tile codes and all other codes, highlighting that improved performance comes from longer-range connections in many code families. In contrast, the tile codes rely on tiling a bigger surface with compact stabilizer tiles. 
        
    \textbf{Bump bonds.} When considering the effect of bump bonds, the variation of hardware complexity across codes seems to be bounded for high-efficiency codes, such as large radial codes. This behavior is expected as HAL allows for a maximum number of bump bonds per edge before aborting the edge and reattempting to route it on a high tier. The maximum number of bump bonds for this data was set to 10. Still, there is an overall increase in hardware complexity with increasing logical efficiency. The weight-4 radial codes perform the best, which can be explained by their lower weight leading to fewer edges, congestion, and crossings, even when compared to the smallest weight-6 tile codes.

    \textbf{TSVs.} In the last row, we single out the contribution from the number of average TSVs per edge. While similar to the first row, where the number of tiers was isolated, the distribution is smoother, capturing the relative fraction of edges in the graph routed on higher tiers. For example, this collapses the cheapest radial and tile codes onto the same point. Though the smallest tile codes have more tiers than the smallest radial codes, their regular structure allows a large fraction of edges to be routed on lower tiers. We also observe that the average number of TSVs increases slightly as the tile codes increase in area. An increasing bulk-to-boundary ratio removes more edges from the congested lower tiers. This code structure reduces the maximum average of bump bonds but increases the utilization of TSVs, explaining the opposite slopes in the weight-6 tile codes when isolating the effect of bump bonds and TSVs.

    \textbf{Resilience to variation in hardware complexity model.} In the second and third column of Fig.~\ref{fig:cost_variation_robustness}, we return to a uniform distribution of weights but change the optimistic fabrication value, \(p_i\), for each of the four parameters by \(\pm50~\%\) from the values in Tab.~\ref{tab:opt_values}. Across all plots, we see that the general trends from Fig.~\ref{fig:complexity} are preserved. The changes caused by variation of \(p_i\) seem to be global shifts and factors across all codes, with minor changes in the relative comparison across codes. A small exception is the improved relative performance of tile codes against other codes when increasing the penalty on average edge length. This trend highlights the greater reliance of many qLDPC codes on long-range couplers than tile codes.

    The minor relative changes in the results plot with up to \(50\%\) changes in the assumptions to our hardware complexity model show that the structure and conclusions we have extracted from our data are robust to uncertainty in our model. Our framework reveals consistent hardware complexity trends across code families. 

    \section{Exploiting Geometric Structure in Bivariate Bicycle Codes} \label{App:exploit_struct}

    The place-and-route workflow in HAL can handle arbitrary codes relying on general heuristics by using a force-directed spring layout to minimize edge lengths regardless of the underlying geometric structure. With more specific knowledge of the code connectivity, however, one can exploit known symmetries to improve the obtained layout. 

    As a case study, we investigate how much it helps to take advantage of the geometric structure in BB codes. The stabilizer structure of the codes in Ref.~\cite{liang2025generalized} ensures the existence of a nearest-neighbor lattice that can be laid out on a square, weight-4 lattice. We can enforce the qubit tier of a code layout to have a square, nearest-neighbor connectivity by explicitly defining the positions of the nodes accordingly---a user-configurable parameter passed as an input to HAL (see App.~\ref{user_configurable_settings}). 

    We lay out all BB codes using both the square grid positions and the generic spring layout in HAL. A clear trend emerges when ordering the resulting hardware complexities by the aspect ratio (AR) of the code, which is given by the ratio of the height of the qubit lattice to its width, and the lattice is oriented such that the height is greater than the width. 

    \begin{figure}[t]
        \centering
        \includegraphics{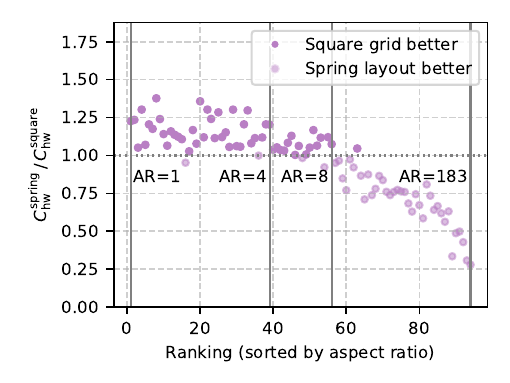}
        \caption{Hardware complexity of square grid and spring layouts for BB codes with varying aspect ratio (AR). Low-aspect-ratio codes show best performance with square grid layouts, while high-aspect-ratio codes have significantly lower hardware complexity using a generic spring layout in HAL.}
        \label{fig:aspect_ratio}
    \end{figure}
    
    The results are shown in Fig.~\ref{fig:aspect_ratio}. Here, we plot the ratio of the hardware complexity obtained by the generic spring layout to the custom square grid for the same code. The codes are ordered by their ranking in aspect ratio, with selective vertical markers indicating the aspect ratio at certain positions. 

    Three distinct regimes are visible in the data: In the low-aspect-ratio regime (AR below 4), the square grid outperforms the spring layout, with the latter being up to \(\sim30\%\) more expensive. In the cross-over regime (AR between 4 and 8), the fraction of higher-performing spring layouts increases slightly, while the square grid still yields the lowest hardware complexity for most codes. In the high-aspect-ratio regime (AR above 8), the spring layout strongly outperforms the square grid, reducing hardware complexities by up to a factor of 4.

    These results illustrate that knowledge of the geometric structure in a code can be exploited to lower the hardware complexity. The structure can be encoded into HAL by explicitly defining the node positions. Still, the advantage of enforcing a square, nearest-neighbor lattice vanishes in the regime of high-aspect-ratio codes. We attribute the diminishing improvement factor to the fact that enforcing a square grid creates long edges in high-aspect ratio, narrow codes. With a reduced bulk size, these edges need to cross many other edges, leading to increased bump bond utilization, tiers, and TSVs, driving up the hardware complexity. 
    
    In regimes where the advantage of exploiting the geometric structure of a code is reduced, the default spring-layout node placement strategy in HAL achieves good performance. This performance underlines the ability of HAL to handle generic code connectivities without a priori knowledge of geometric structure.

    \section{Varying Check Qubit Positions\\in Tile Codes}\label{App:vary_check_qubit_pos}

    Following the construction of tile codes introduced in Ref.~\cite{steffan2025tile}, stabilizers are defined on tiles, which are grids of bounded area, where the grid edges represent data qubits. We adapt a figure for the stabilizers of the \(\llbracket288,8,14\rrbracket\) code from Ref.~\cite{steffan2025tile}:

    \begin{figure}[htbp]
        \centering
        \includegraphics[]{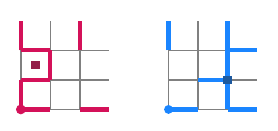}
    \end{figure}

    The red edges represent the data qubits supported in this code's X-stabilizer, and the blue edges the data qubits in the Z-stabilizer. These tiles are then repeated across a plane to define the code. At the boundaries, the stabilizers are truncated. Unchecked data qubits are removed, and so are any resulting empty stabilizers. See \cite{steffan2025tile} for more details.

    This construction already provides the connectivity graph---we assume one check qubit for each tile. However, to enforce the regularity possible with tile codes, we need to fix the position of the check qubits and provide all qubit positions as a geometric ansatz to HAL. We allow for check qubits of one basis to be placed on faces of the grid and check qubits of the other basis on vertices, such that each position is occupied at most once as the plane is tiled. Example positions are highlighted with squares in the example tiles. 

    Since each check qubit has a fixed edge to all data qubits in its support, the check qubit position strongly impacts the layout of these edges and, thus, the total hardware complexity of the code. We define different heuristics for choosing check qubit positions and study their impact on hardware complexity when applied to three different example codes (see Tab.~\ref{tab:tile_code_heuristics}). The heuristics are defined as follows:
    \begin{itemize}
        \item \textbf{random}: Choose a random position.
        \item \textbf{manhattan}: Minimize sum of Manhattan distances to data qubits.
        \item \textbf{euclidean}: Minimize sum of Euclidean distances to data qubits.
        \item \textbf{nearest-neighbor}: Maximize number of nearest-neighbor connections to data qubits. 
        \item \textbf{manual}: Manually choose a position.
    \end{itemize}

    We lay out ten instances of \emph{random} for all three codes and extract the average hardware complexity and its standard deviation. The other heuristics are deterministic, so we only produce one layout, respectively. Across all codes, \emph{random} consistently returns hardware complexities that are an average of around 16-19\% higher than the best heuristic. The overall best performing heuristic is \emph{euclidean}, which is just slightly outperformed once by \emph{manual}. While \emph{nearest-neighbor} can achieve good performance, it leads to a 16\% higher hardware complexity than the optimal for the weight-10 code.

    \begin{table}[htbp]
    \centering
    \begin{tabularx}{\linewidth}{l@{\hskip 0.7cm}l@{\hskip 0.8cm}X}
    \toprule
    Code & Heuristic & \arraybackslash $C_{hw}$ \\
    \midrule
    \multirow{4}{*}{\shortstack{\(\llbracket188,8,9\rrbracket\)\\($w=6$)}}
      & random            & \(1.799 \pm 0.066\) \\
      & manhattan         & \(1.618\) \\
      & euclidean         & \(1.541\) \\
      & nearest\_neighbor & \(1.541\) \\
    \midrule
    \multirow{5}{*}{\shortstack{\(\llbracket292,12,14\rrbracket\)\\($w=8$)}}
      & random            & \(2.022 \pm 0.128\) \\
      & manhattan         & \(1.833\) \\
      & euclidean         & \(1.759\) \\
      & nearest\_neighbor & \(1.833\) \\
      & manual            & \(1.696\) \\
    \midrule
    \multirow{4}{*}{\shortstack{\(\llbracket512,18,23\rrbracket\)\\($w=10$)}}
      & random            & \(2.335 \pm 0.218\) \\
      & manhattan         & \(1.958\) \\
      & euclidean         & \(1.958\) \\
      & nearest\_neighbor & \(2.278\) \\
    \bottomrule
    \end{tabularx}
    \caption{Hardware complexity (mean $\pm$ std) for different codes and check-qubit positioning heuristics.}
    \label{tab:tile_code_heuristics}
    \end{table}

    We provide some intuition for these results. Random position assignments can lead to long edges, which also entails more crossings and the need for more tiers. This can vary greatly depending on the check qubit position, as evidenced by the standard deviation. All other heuristics attempt to reduce total edge length by different metrics, which seems to lead to lower hardware complexities.
    
    The \emph{nearest-neighbor} heuristic maximizes the number of edges that can be placed on the qubit tier, where edges are not allowed to cross. However, this greedy approach often yields longer edges on higher tiers, which leads to more crossings. This is especially detrimental to higher weight codes. Since the qubit tier supports at most a qubit degree of four, high-weight codes inevitably place many edges on higher tiers, where this heuristic performs poorly. As seen in Fig.~\ref{tab:tile_code_heuristics}, the hardware complexity increases with higher code weight.

    The \emph{euclidean} strategy performs consistently well, except for one case where a slightly better manual placement was found. Its effectiveness stems from the fact that most tile-code edges can be routed as straight lines, so the total Euclidean distance between a check and its data qubits is a good proxy for edge length and crossings. Unlike the more greedy \emph{nearest-neighbor} heuristic, it does not overemphasize the qubit tier and thus remains robust as stabilizer weight increases.

    In practice, we choose the heuristic that yields the lowest hardware complexity for each tile code. Because of the regularity of tile codes, an optimal heuristic for one code should remain optimal for a larger code that uses the same underlying tiles. We can, therefore, probe the optimal heuristic using smaller instances of a tile code to speed up the optimization.
    
    \section{Distance Estimation of Radial Codes} \label{app:dist_est_radial}

    Our analysis of quantum radial codes closely follows Ref.~\cite{scruby2024high}. These codes are generated from a lifted product of two random classical radial codes with $r$ concentric rings and $s$ spokes. As such, there are many possible code instances for a given $(r,s)$-pair that have the code parameters $\llbracket 2r^2s,2(r-1)^2, \leq2s\rrbracket$. Note that the distance is only provided as an upper bound. While it is believed that a code instance exists for every $(r,s)$-pair that saturates this bound, i.e., $d=2s$, there is no known method to reliably achieve this upper bound.\par
    In this work, we rely on the package \texttt{QDistRnd} \cite{pryadko2023qdistrnd} to numerically estimate the distances of random code instances. We improve the efficiency of our search for high-distance radial codes through the use of a batch-decoding and distance-pruning strategy. For a given $(r,s)$-pair, we initialize a variable to track the estimated distance, $d_\text{est}=0$. We generate batches of 125 random code instances, which we pass into a single \texttt{GAP} subroutine. For each instance in the batch, we run \texttt{DistRandCSS}. As soon as a code instance is found to have a distance less than or equal to $d_\text{est}$, the algorithm aborts and moves on to the next code instance. If the distance of a code instance does not drop below the estimated distance, we update $d_\text{est}$ to the newly found value, and the algorithm proceeds with the next code instance. The maximum number of trials for each call of \texttt{DistRandCSS} is set to $1e6$, which was shown to yield high-confidence estimates of the distance in Ref.~\cite{scruby2024high}.\par

    \begin{table}[htbp]
    \centering
    \small
    \begin{tabularx}{\columnwidth}{l @{\hskip 25pt} l @{\hskip 25pt} X}
        \toprule
        \multirow{2}{*}{$(r,s)$} &
        \multirow{2}{*}{Distances $d_\text{est}$} &
        \multirow{2}{*}{$C^{d<d_\text{max}}_{\text{hw,min}}/C^{d=d_\text{max}}_\text{hw,min}$} \\ \\
        \midrule
        \textbf{(2,2)}  & $\{\textbf{4}\}$ & -- \\
        (2,3)  & $\{4\}$ & -- \\
        (2,5)  & $\{4,6\}$ & 1.051 \\
        (2,7)  & $\{4,6\}$ & 1.008 \\
        (2,11) & $\{4,6,8\}$ & 0.979 \\
        (2,13) & $\{4,6,8,10\}$ & 0.982 \\
        \midrule
        \textbf{(3,3)}  & $\{\textbf{6}\}$ & -- \\
        \textbf{(3,5)}  & $\{6,8,\textbf{10}\}$ & 1.005 \\
        (3,7)  & $\{6,8,10\}$ & 0.993 \\
        (3,11) & $\{8,10,12,14\}$ & 1.000 \\
        (3,13) & $\{8,10,12,14\}$ & 0.969 \\
        \midrule
        (4,5)  & $\{8\}$ & -- \\
        \textbf{(4,7)} & $\{8,10,12,\textbf{14}\}$ & 0.984 \\
        \textbf{(4,11)} & $\{16,18,20,\textbf{22}\}$ & 1.007 \\
        \textbf{(4,13)} & $\{16,18,20,22,24,\textbf{26}\}$ & 0.993 \\
        \bottomrule
    \end{tabularx}
    \caption{Obtained distances of radial code instances for different values of $(r,s)$, verified by \texttt{QDistRnd} \cite{pryadko2023qdistrnd}. For each $(r,s)$, we take the ratio of the minimum hardware complexity for instances with distance $d<d_\text{max}$ and with $d=d_\text{max}$, where $d_\text{max}$ is the maximum obtained distance for that $(r,s)$. Radial codes $(r,s)$ with instances that saturate the distance bound $d=2s$ are highlighted in bold.}
    \label{tab:radial_code_distances}
    \end{table}

    Table \ref{tab:radial_code_distances} shows the values of $(r,s)$ for radial codes investigated in our work. The middle column lists the distances found after sampling $10,000$ instances. Codes for which we found instances that saturate the distance bound, $d=2s$, are highlighted in bold. \par
    Furthermore, we investigate whether radial codes of the same $(r,s)$ but greater distance have a higher hardware complexity. To this end, we use HAL to lay out ten code instances of the same $(r,s)$, uniformly distributed across the obtained distances. We also lay out one code instance of the highest distance ten times. Since some of the heuristics HAL uses are stochastic, the repeated layout of the same code instance can vary. We then compute the ratio between the minimum hardware complexity across codes with distance smaller than the highest obtained distance, $d_\text{max}$, which we denote as $C^{d<d_\text{max}}_{\mathrm{hw,min}}$, and across the layouts of the same code instance, $C^{d=d_\text{max}}_\text{hw,min}$ (shown in the third column of Tab.~\ref{tab:radial_code_distances}). A ratio smaller than 1 reflects that code instances with a lower distance have a lower hardware complexity. This metric is redundant for codes where we only found instances of the same distance. \par
    We observe that the hardware complexity of radial code instances reaching a high distance does not differ significantly from that of instances with low distances, since the complexity ratios in Tab.~\ref{app:dist_est_radial} are close to 1 across all radial codes. From Fig.~\ref{fig:complexity}, the dominant mechanisms driving the hardware complexity in radial codes are the values for $(r,s)$, which set the number of qubits, $n=2r^2s$, and the weight of the code, $w=2r$. This is in line with the findings for tile codes, where the main contribution to the hardware complexity is the weight of the code, with a small dependence on the number of qubits.\par
    Since the realized distance of a radial code instance seems independent of the hardware complexity, any code instance for a value of $(r,s)$ can be used as a proxy to estimate the hardware complexity of radial codes with a higher distance for the same $(r,s)$. Furthermore, if it is possible to reliably find a code instance for any $(r,s)$ that saturates the upper distance bound, we can use a lower-distance proxy code to estimate the hardware complexity of a corresponding radial code with distance $d=2s$. The results on radial codes presented in our work follow this approach.\par
    Our investigation points to an important research avenue for future work---understanding the distance-reducing mechanism in lifted product codes. Among the most promising codes identified in Sec.~\ref{sec:results} are weight-4 radial codes ($r=2$), of which, however, we were unable to reliably find code instances that saturate the distance bound. Yet, there is reason to believe that a code instance exists for every $(r,s)$ with distance $d=2s$, saturating the upper bound. First, the quantum radial code construction involves the lifted product of two classical radial codes that both have a verifiable distance of $d=2s$. Second, for many values of $(r,s)$, it is indeed possible to find codes that reach the highest possible distance, distributed across a wide range of values for $(r,s)$. Third, a distance-reducing mechanism studied in \cite{scruby2024high} is a coincidental consequence of the lifted product involving random codes. With certain combinations of classical input codes, low-weight logicals do not appear in the quantum code, preventing the distance from falling below the distance of the classical codes.\par
    Understanding why the lifted product often results in drastic reductions of the maximally attainable distance is an open problem. This stands in contrast to the hypergraph product, in which the distances of the classical input codes carry over to the quantum code in a straightforward manner \cite{breuckmann2021quantum}. Developing a technique to reliably prevent the lifted product code from reducing the distance could unlock highly promising codes that achieve excellent tradeoffs between logical efficiency and hardware complexity for superconducting qubits.

    \section{Database of Code Layouts}
    
    Table~\ref{tab:metrics_database} contains all 144 codes laid out in this work and shown in Fig.~\ref{fig:complexity}. It shows the code parameters, logical efficiency, individual hardware parameters, and final hardware complexity for each QECC. Furthermore, the layouts for all the codes in Fig.~\ref{fig:complexity} can be visualized using an online database at \cite{hal_database}.
    
    \begin{table*}[htbp]
    \centering
    \renewcommand{\arraystretch}{1.4}
    \small
    \newcolumntype{Y}{>{\raggedright\arraybackslash}m{0.48\textwidth}}
    \begin{tabularx}{\textwidth}{@{}Y@{\hspace{1em}}Y@{}}
      \begin{minipage}[t]{\linewidth}\label{tab:full_database}
        \vspace{0pt}
        \centering
        \setlength{\tabcolsep}{4pt}
        \begin{tabular}{@{}lc|cccc|c@{}}
          \toprule
          \(\llbracket n,k,d\rrbracket\) & \(kd^2/n\) & Tiers & Length & Bumps & TSVs & \(C_\text{hw}\) \\
          \midrule 
          \multicolumn{3}{c}{\textbf{Surface code}} & \multicolumn{2}{c}{Total: 1} & \multicolumn{2}{c}{Ref.: \cite{acharya2024quantum}}
          \\ \midrule
          \(\llbracket n,1,\sqrt{n}\rrbracket\) & 1 & 1 & 1.00 & 0.00 & 0.00 & 1.00 \\
          \midrule 
          \multicolumn{3}{c}{\textbf{Directional codes}} & \multicolumn{2}{c}{Total: 10}& \multicolumn{2}{c}{Ref.: \cite{geher2025directional}}
          \\ \midrule
          \(\llbracket144,6,6\rrbracket\) & 1.5 & 2 & 8.39 & 2.13 & 2.00 & 1.57 \\
          \(\llbracket64,6,4\rrbracket\) & 1.5 & 2 & 5.79 & 2.19 & 2.00 & 1.50 \\
          \(\llbracket256,6,8\rrbracket\) & 1.5 & 2 & 10.94 & 2.10 & 2.00 & 1.64 \\
          \(\llbracket36,4,4\rrbracket\) & 1.78 & 2 & 3.43 & 2.86 & 2.00 & 1.48 \\
          \(\llbracket72,4,6\rrbracket\) & 2.0 & 2 & 5.94 & 2.75 & 2.00 & 1.54 \\
          \(\llbracket120,4,8\rrbracket\) & 2.13 & 2 & 6.13 & 2.79 & 2.00 & 1.55 \\
          \(\llbracket180,4,10\rrbracket\) & 2.22 & 2 & 8.74 & 2.73 & 2.00 & 1.62 \\
          \(\llbracket288,12,8\rrbracket\) & 2.67 & 2 & 15.18 & 1.99 & 2.00 & 1.75 \\
          \(\llbracket144,12,6\rrbracket\) & 3.0 & 2 & 10.58 & 2.06 & 2.00 & 1.62 \\
          \(\llbracket48,12,4\rrbracket\) & 4.0 & 2 & 6.07 & 2.25 & 2.00 & 1.51 \\
          \midrule 
          \multicolumn{3}{c}{\textbf{BB codes}} & \multicolumn{2}{c}{Total: 49}& \multicolumn{2}{c}{Ref.: \cite{liang2025generalized}}
          \\ \midrule
          \(\llbracket24,4,4\rrbracket\) & 2.67 & 3 & 3.65 & 2.50 & 2.13 & 1.53 \\
          \(\llbracket28,6,4\rrbracket\) & 3.43 & 3 & 4.74 & 4.35 & 2.30 & 1.69 \\
          \(\llbracket18,4,4\rrbracket\) & 3.56 & 3 & 3.07 & 3.04 & 2.12 & 1.55 \\
          \(\llbracket30,4,6\rrbracket\) & 4.8 & 3 & 3.74 & 3.71 & 2.21 & 1.62 \\
          \(\llbracket60,8,6\rrbracket\) & 4.8 & 4 & 8.80 & 3.36 & 2.64 & 1.83 \\
          \(\llbracket78,4,10\rrbracket\) & 5.13 & 5 & 11.23 & 3.25 & 3.09 & 1.99 \\
          \(\llbracket42,6,6\rrbracket\) & 5.14 & 3 & 5.25 & 3.95 & 2.37 & 1.69 \\
          \(\llbracket48,4,8\rrbracket\) & 5.33 & 4 & 7.17 & 3.83 & 2.44 & 1.80 \\
          \(\llbracket54,8,6\rrbracket\) & 5.33 & 5 & 7.29 & 3.50 & 2.86 & 1.88 \\
          \(\llbracket70,6,8\rrbracket\) & 5.49 & 4 & 6.85 & 3.17 & 2.38 & 1.75 \\
          \(\llbracket102,4,12\rrbracket\) & 5.65 & 5 & 14.49 & 3.89 & 3.09 & 2.13 \\
          \(\llbracket138,4,14\rrbracket\) & 5.68 & 5 & 13.48 & 3.93 & 3.19 & 2.11 \\
          \(\llbracket96,4,12\rrbracket\) & 6.0 & 5 & 9.61 & 5.08 & 3.06 & 2.06 \\
          \(\llbracket66,4,10\rrbracket\) & 6.06 & 4 & 8.78 & 2.33 & 2.64 & 1.77 \\
          \(\llbracket56,6,8\rrbracket\) & 6.86 & 4 & 6.39 & 3.64 & 2.69 & 1.79 \\
          \(\llbracket84,6,10\rrbracket\) & 7.14 & 4 & 10.84 & 4.13 & 3.14 & 1.98 \\
          \(\llbracket108,8,10\rrbracket\) & 7.41 & 5 & 8.92 & 5.18 & 3.33 & 2.07 \\
          \(\llbracket258,4,22\rrbracket\) & 7.5 & 6 & 19.91 & 4.69 & 5.34 & 2.58 \\
          \(\llbracket150,8,12\rrbracket\) & 7.68 & 5 & 19.02 & 4.31 & 3.39 & 2.30 \\
          \(\llbracket112,6,12\rrbracket\) & 7.71 & 4 & 8.70 & 4.42 & 2.99 & 1.93 \\
          \(\llbracket288,16,12\rrbracket\) & 8.0 & 6 & 15.73 & 5.00 & 3.51 & 2.33 \\
          \(\llbracket276,4,24\rrbracket\) & 8.35 & 6 & 19.29 & 3.90 & 3.50 & 2.36 \\
          \(\llbracket140,6,14\rrbracket\) & 8.4 & 5 & 10.47 & 3.66 & 2.99 & 1.99 \\
          \(\llbracket98,6,12\rrbracket\) & 8.82 & 5 & 8.49 & 5.86 & 3.11 & 2.08 \\
          \(\llbracket170,16,10\rrbracket\) & 9.41 & 6 & 13.13 & 5.18 & 3.30 & 2.25 \\
          \(\llbracket126,12,10\rrbracket\) & 9.52 & 5 & 9.91 & 5.40 & 2.91 & 2.08 \\
          \bottomrule
        \end{tabular}
      \end{minipage} &
      \begin{minipage}[t]{\linewidth}
        \vspace{0pt}
        \centering
        \setlength{\tabcolsep}{4pt}
        \begin{tabular}{@{}lc|cccc|c@{}}
          \toprule
          \(\llbracket n,k,d\rrbracket\) & \(kd^2/n\) & Tiers & Length & Bumps & TSVs & \(C_\text{hw}\) \\
          \midrule
          \(\llbracket120,8,12\rrbracket\) & 9.6 & 4 & 9.72 & 4.83 & 2.57 & 1.95 \\
          \(\llbracket162,8,14\rrbracket\) & 9.68 & 5 & 10.69 & 4.82 & 3.53 & 2.12 \\
          \(\llbracket238,6,20\rrbracket\) & 10.08 & 6 & 14.55 & 4.51 & 3.82 & 2.29 \\
          \(\llbracket280,6,22\rrbracket\) & 10.37 & 6 & 25.78 & 4.42 & 3.33 & 2.56 \\
          \(\llbracket192,8,16\rrbracket\) & 10.67 & 6 & 12.84 & 4.85 & 3.66 & 2.25 \\
          \(\llbracket182,6,18\rrbracket\) & 10.68 & 4 & 10.34 & 4.26 & 3.17 & 1.98 \\
          \(\llbracket224,6,20\rrbracket\) & 10.71 & 5 & 20.96 & 3.72 & 3.15 & 2.30 \\
          \(\llbracket322,6,24\rrbracket\) & 10.73 & 6 & 18.66 & 4.84 & 3.85 & 2.43 \\
          \(\llbracket240,8,18\rrbracket\) & 10.8 & 5 & 13.78 & 5.39 & 2.81 & 2.18 \\
          \(\llbracket266,6,22\rrbracket\) & 10.92 & 5 & 17.51 & 4.25 & 2.88 & 2.21 \\
          \(\llbracket364,6,26\rrbracket\) & 11.14 & 7 & 17.45 & 5.62 & 3.70 & 2.49 \\
          \(\llbracket180,8,16\rrbracket\) & 11.38 & 6 & 13.55 & 5.63 & 3.95 & 2.34 \\
          \(\llbracket350,6,26\rrbracket\) & 11.59 & 7 & 26.45 & 5.04 & 4.08 & 2.74 \\
          \(\llbracket324,8,22\rrbracket\) & 11.95 & 6 & 15.23 & 4.35 & 3.89 & 2.30 \\
          \(\llbracket392,6,28\rrbracket\) & 12.0 & 7 & 22.32 & 5.16 & 4.35 & 2.65 \\
          \(\llbracket252,12,16\rrbracket\) & 12.19 & 6 & 17.20 & 5.01 & 3.64 & 2.38 \\
          \(\llbracket210,10,16\rrbracket\) & 12.19 & 7 & 16.96 & 5.79 & 5.35 & 2.63 \\
          \(\llbracket294,10,20\rrbracket\) & 13.61 & 6 & 20.20 & 4.77 & 3.64 & 2.45 \\
          \(\llbracket390,8,26\rrbracket\) & 13.87 & 6 & 18.42 & 5.09 & 3.35 & 2.39 \\
          \(\llbracket340,16,18\rrbracket\) & 15.25 & 7 & 28.40 & 4.31 & 3.76 & 2.72 \\
          \(\llbracket378,12,22\rrbracket\) & 15.37 & 7 & 20.09 & 5.16 & 4.09 & 2.57 \\
          \(\llbracket310,10,22\rrbracket\) & 15.61 & 8 & 22.81 & 5.82 & 5.64 & 2.88 \\
          \(\llbracket360,12,24\rrbracket\) & 19.2 & 8 & 20.46 & 5.32 & 4.70 & 2.70 \\
          \midrule \multicolumn{3}{c}{\textbf{Narrow BB codes}} & \multicolumn{2}{c}{Total: 43}& \multicolumn{2}{c}{Ref.: \cite{liang2025generalized}} \\ \midrule
          \(\llbracket12,4,2\rrbracket\) & 1.33 & 3 & 3.77 & 2.38 & 2.16 & 1.53 \\
          \(\llbracket14,6,2\rrbracket\) & 1.71 & 3 & 5.66 & 3.02 & 2.08 & 1.62 \\
          \(\llbracket146,18,4\rrbracket\) & 1.97 & 6 & 16.25 & 4.97 & 3.09 & 2.30 \\
          \(\llbracket292,18,8\rrbracket\) & 3.95 & 9 & 18.73 & 6.17 & 4.37 & 2.74 \\
          \(\llbracket36,4,6\rrbracket\) & 4.0 & 3 & 6.65 & 3.31 & 2.22 & 1.67 \\
          \(\llbracket62,10,6\rrbracket\) & 5.81 & 5 & 10.09 & 4.80 & 2.87 & 2.04 \\
          \(\llbracket132,4,14\rrbracket\) & 5.94 & 6 & 13.32 & 5.40 & 3.34 & 2.27 \\
          \(\llbracket156,4,16\rrbracket\) & 6.56 & 7 & 14.26 & 5.55 & 3.56 & 2.39 \\
          \(\llbracket114,4,14\rrbracket\) & 6.88 & 6 & 13.18 & 5.46 & 3.37 & 2.27 \\
          \(\llbracket228,4,20\rrbracket\) & 7.02 & 7 & 19.88 & 5.96 & 3.80 & 2.59 \\
          \(\llbracket72,8,8\rrbracket\) & 7.11 & 5 & 8.29 & 5.42 & 2.98 & 2.04 \\
          \(\llbracket222,4,20\rrbracket\) & 7.21 & 7 & 19.43 & 6.21 & 3.80 & 2.59 \\
          \(\llbracket174,4,18\rrbracket\) & 7.45 & 9 & 16.78 & 5.59 & 4.52 & 2.66 \\
          \(\llbracket348,4,26\rrbracket\) & 7.77 & 8 & 20.81 & 5.99 & 4.31 & 2.72 \\
          \(\llbracket204,4,20\rrbracket\) & 7.84 & 7 & 16.45 & 5.80 & 4.07 & 2.51 \\
          \(\llbracket246,4,22\rrbracket\) & 7.87 & 8 & 21.76 & 5.86 & 4.16 & 2.73 \\
          \(\llbracket124,10,10\rrbracket\) & 8.06 & 7 & 11.78 & 5.90 & 3.82 & 2.36 \\
          \bottomrule
        \end{tabular}
      \end{minipage}
    \end{tabularx}
    \caption{\textbf{Table of all codes laid out in Fig.~\ref{fig:complexity}.} The codes are ordered by increasing logical efficiency within each code family. The four individual hardware parameters represent the raw quantities extracted from the final layout. The parameters are rescaled and combined into a single hardware complexity \(C_\text{hw}\), following the model introduced in Sec.~\ref{sec:results}. All layouts can be viewed at \cite{hal_database}.}
    \label{tab:metrics_database}
    \end{table*}
    
    \begin{table*}[htbp]
    \centering
    \renewcommand{\arraystretch}{1.4}
    \small
    \newcolumntype{Y}{>{\raggedright\arraybackslash}m{0.48\textwidth}}
    \begin{tabularx}{\textwidth}{@{}Y@{\hspace{1em}}Y@{}}
      \begin{minipage}[t]{\linewidth}
        \vspace{0pt}
        \centering
        \setlength{\tabcolsep}{4pt}
        \begin{tabular}{@{}lc|cccc|c@{}}
          \toprule
          \(\llbracket n,k,d\rrbracket\) & \(kd^2/n\) & Tiers & Length & Bumps & TSVs & \(C_\text{hw}\) \\
          \midrule
          \(\llbracket282,4,24\rrbracket\) & 8.17 & 8 & 19.69 & 5.97 & 3.94 & 2.66 \\
          \(\llbracket318,4,26\rrbracket\) & 8.5 & 9 & 23.94 & 5.89 & 4.87 & 2.91 \\
          \(\llbracket366,4,28\rrbracket\) & 8.57 & 10 & 25.01 & 6.23 & 5.15 & 3.05 \\
          \(\llbracket354,4,28\rrbracket\) & 8.86 & 9 & 20.08 & 5.95 & 4.65 & 2.79 \\
          \(\llbracket90,8,10\rrbracket\) & 8.89 & 6 & 11.34 & 5.27 & 3.33 & 2.21 \\
          \(\llbracket168,8,14\rrbracket\) & 9.33 & 7 & 17.75 & 5.48 & 3.49 & 2.47 \\
          \(\llbracket196,6,18\rrbracket\) & 9.92 & 7 & 18.80 & 5.94 & 3.77 & 2.55 \\
          \(\llbracket154,6,16\rrbracket\) & 9.97 & 6 & 14.13 & 5.44 & 3.50 & 2.31 \\
          \(\llbracket198,8,16\rrbracket\) & 10.34 & 8 & 17.51 & 6.20 & 3.88 & 2.61 \\
          \(\llbracket186,10,14\rrbracket\) & 10.54 & 7 & 17.61 & 5.63 & 3.76 & 2.50 \\
          \(\llbracket234,8,18\rrbracket\) & 11.08 & 8 & 17.90 & 5.92 & 4.09 & 2.62 \\
          \(\llbracket308,6,24\rrbracket\) & 11.22 & 9 & 17.07 & 5.84 & 4.67 & 2.70 \\
          \(\llbracket342,8,22\rrbracket\) & 11.32 & 11 & 26.93 & 6.11 & 5.37 & 3.17 \\
          \(\llbracket270,8,20\rrbracket\) & 11.85 & 11 & 19.85 & 5.84 & 4.78 & 2.91 \\
          \(\llbracket216,8,18\rrbracket\) & 12.0 & 8 & 18.53 & 6.02 & 3.69 & 2.61 \\
          \(\llbracket264,8,20\rrbracket\) & 12.12 & 8 & 19.23 & 5.75 & 4.19 & 2.65 \\
          \(\llbracket312,8,22\rrbracket\) & 12.41 & 9 & 21.71 & 5.71 & 4.18 & 2.78 \\
          \(\llbracket306,8,22\rrbracket\) & 12.65 & 9 & 19.23 & 6.11 & 4.69 & 2.78 \\
          \(\llbracket300,8,22\rrbracket\) & 12.91 & 8 & 18.72 & 6.09 & 4.24 & 2.66 \\
          \(\llbracket248,10,18\rrbracket\) & 13.06 & 8 & 18.14 & 5.96 & 4.18 & 2.63 \\
          \(\llbracket396,8,26\rrbracket\) & 13.66 & 10 & 23.66 & 6.32 & 5.49 & 3.04 \\
          \(\llbracket330,8,24\rrbracket\) & 13.96 & 10 & 19.46 & 5.79 & 4.82 & 2.84 \\
          \(\llbracket254,14,16\rrbracket\) & 14.11 & 8 & 19.81 & 6.40 & 4.45 & 2.73 \\
          \(\llbracket336,10,22\rrbracket\) & 14.4 & 9 & 21.71 & 6.34 & 4.99 & 2.89 \\
          \(\llbracket372,10,24\rrbracket\) & 15.48 & 10 & 22.20 & 6.04 & 5.01 & 2.95 \\
          \(\llbracket384,12,24\rrbracket\) & 18.0 & 11 & 23.46 & 6.44 & 5.32 & 3.09 \\
          \midrule \multicolumn{3}{c}{\textbf{Gross codes}} & \multicolumn{2}{c}{Total: 2}& \multicolumn{2}{c}{Ref.: \cite{bravyi2024high}}\\ \midrule
          \(\llbracket144,12,12\rrbracket\) & 12.0 & 5 & 11.08 & 5.06 & 3.27 & 2.12 \\
          \(\llbracket288,12,18\rrbracket\) & 13.5 & 5 & 13.94 & 5.13 & 3.75 & 2.24 \\
          \midrule \multicolumn{3}{c}{\textbf{Tile codes}} & \multicolumn{2}{c}{Total: 19}& \multicolumn{2}{c}{Ref.: \cite{steffan2025tile,liang2025planar}} \\ \midrule
          \(\llbracket105,8,6\rrbracket\) & 2.74 & 3 & 2.91 & 2.96 & 2.14 & 1.54 \\
          \(\llbracket137,8,7\rrbracket\) & 2.86 & 3 & 2.89 & 2.98 & 2.13 & 1.54 \\
          \(\llbracket173,8,8\rrbracket\) & 2.96 & 3 & 2.96 & 2.90 & 2.15 & 1.54 \\
          \(\llbracket188,8,9\rrbracket\) & 3.45 & 3 & 2.98 & 2.89 & 2.17 & 1.54 \\
          \(\llbracket230,8,10\rrbracket\) & 3.48 & 3 & 2.87 & 2.84 & 2.18 & 1.54 \\
          \(\llbracket276,8,11\rrbracket\) & 3.51 & 3 & 2.93 & 2.87 & 2.21 & 1.54 \\
          \(\llbracket295,8,12\rrbracket\) & 3.91 & 3 & 2.85 & 2.86 & 2.21 & 1.54 \\
          \(\llbracket326,8,13\rrbracket\) & 4.15 & 3 & 2.86 & 2.62 & 2.28 & 1.53 \\
          \(\llbracket347,8,14\rrbracket\) & 4.52 & 3 & 2.97 & 2.72 & 2.26 & 1.54 \\
          \(\llbracket368,8,15\rrbracket\) & 4.89 & 3 & 2.88 & 2.63 & 2.30 & 1.53 \\
          \bottomrule
        \end{tabular}
      \end{minipage} &
      \begin{minipage}[t]{\linewidth}
        \vspace{0pt}
        \centering
        \setlength{\tabcolsep}{4pt}
        \begin{tabular}{@{}lc|cccc|c@{}}
          \toprule
          \(\llbracket n,k,d\rrbracket\) & \(kd^2/n\) & Tiers & Length & Bumps & TSVs & \(C_\text{hw}\) \\
          \midrule
          \(\llbracket227,12,11\rrbracket\) & 6.4 & 5 & 4.12 & 3.26 & 2.62 & 1.76 \\
          \(\llbracket240,12,12\rrbracket\) & 7.2 & 5 & 3.95 & 3.17 & 2.66 & 1.75 \\
          \(\llbracket292,12,14\rrbracket\) & 8.05 & 5 & 4.19 & 3.18 & 2.77 & 1.77 \\
          \(\llbracket365,12,16\rrbracket\) & 8.42 & 5 & 4.36 & 2.86 & 2.96 & 1.77 \\
          \(\llbracket382,12,17\rrbracket\) & 9.08 & 5 & 4.25 & 2.86 & 2.86 & 1.76 \\
          \(\llbracket399,12,18\rrbracket\) & 9.74 & 5 & 4.11 & 2.88 & 2.95 & 1.76 \\
          \(\llbracket288,18,13\rrbracket\) & 10.56 & 5 & 3.24 & 3.26 & 3.16 & 1.78 \\
          \(\llbracket512,18,19\rrbracket\) & 12.69 & 6 & 3.36 & 2.75 & 3.64 & 1.85 \\
          \(\llbracket512,18,23\rrbracket\) & 18.6 & 7 & 3.73 & 2.69 & 4.08 & 1.96 \\
          \midrule \multicolumn{3}{c}{\textbf{Radial codes}} & \multicolumn{2}{c}{Total: 15}& \multicolumn{2}{c}{Ref.: \cite{scruby2024high}}\\ \midrule
          \(\llbracket16,2,4\rrbracket\) & 2.0 & 2 & 2.76 & 1.78 & 2.00 & 1.39 \\
          \(\llbracket24,2,6\rrbracket\) & 3.0 & 2 & 3.28 & 2.36 & 2.00 & 1.44 \\
          \(\llbracket40,2,10\rrbracket\) & 5.0 & 2 & 4.92 & 2.23 & 2.00 & 1.48 \\
          \(\llbracket54,8,6\rrbracket\) & 5.33 & 4 & 9.18 & 5.18 & 2.68 & 1.96 \\
          \(\llbracket56,2,14\rrbracket\) & 7.0 & 2 & 3.84 & 2.54 & 2.00 & 1.47 \\
          \(\llbracket90,8,10\rrbracket\) & 8.89 & 5 & 12.69 & 5.61 & 2.93 & 2.17 \\
          \(\llbracket88,2,22\rrbracket\) & 11.0 & 3 & 11.05 & 2.45 & 2.07 & 1.73 \\
          \(\llbracket160,18,10\rrbracket\) & 11.25 & 8 & 15.51 & 6.31 & 4.85 & 2.64 \\
          \(\llbracket126,8,14\rrbracket\) & 12.44 & 5 & 13.19 & 5.30 & 3.16 & 2.18 \\
          \(\llbracket104,2,26\rrbracket\) & 13.0 & 3 & 10.00 & 2.59 & 2.16 & 1.72 \\
          \(\llbracket224,18,14\rrbracket\) & 15.75 & 10 & 20.58 & 6.78 & 5.98 & 3.03 \\
          \(\llbracket198,8,22\rrbracket\) & 19.56 & 7 & 17.82 & 5.94 & 3.76 & 2.53 \\
          \(\llbracket234,8,26\rrbracket\) & 23.11 & 8 & 20.88 & 5.95 & 3.99 & 2.69 \\
          \(\llbracket352,18,22\rrbracket\) & 24.75 & 14 & 26.87 & 6.81 & 8.15 & 3.64 \\
          \(\llbracket416,18,26\rrbracket\) & 29.25 & 15 & 33.47 & 6.84 & 8.86 & 3.94 \\
          \midrule \multicolumn{3}{c}{\textbf{Tanner code}} & \multicolumn{2}{c}{Total: 5}& \multicolumn{2}{c}{Ref.: \cite{radebold2025explicit}} \\ \midrule
          \(\llbracket36,8,3\rrbracket\) & 2.0 & 3 & 7.30 & 3.39 & 2.16 & 1.69 \\
          \(\llbracket72,14,4\rrbracket\) & 3.11 & 4 & 8.15 & 4.86 & 2.59 & 1.91 \\
          \(\llbracket54,11,4\rrbracket\) & 3.26 & 4 & 7.77 & 3.89 & 2.40 & 1.82 \\
          \(\llbracket200,10,10\rrbracket\) & 5.0 & 9 & 17.45 & 5.48 & 4.84 & 2.70 \\
          \(\llbracket250,10,15\rrbracket\) & 9.0 & 11 & 21.63 & 5.93 & 5.73 & 3.05 \\
          \bottomrule
        \end{tabular}
        \\[1em]
        \parbox[t]{\linewidth}{\raggedright\hspace{0.0em}TABLE~\thetable \hspace{0.5em}(continued). Table of all codes laid out in Fig.~\ref{fig:complexity}. The codes are ordered by increasing logical efficiency within each code family. The four individual hardware parameters represent the raw quantities extracted from the final layout. The parameters are rescaled and combined into a single hardware complexity \(C_\text{hw}\), following the model introduced in Sec.~\ref{sec:results}. All layouts can be viewed at \cite{hal_database}.}
      \end{minipage}
    \end{tabularx}
    \end{table*}

    \FloatBarrier
    \nocite{*}
    \bibliography{my_bibliography}
\end{document}